\pdfoutput=1
\documentclass[
 aip,
 amsmath,amssymb,
 reprint,
 floatfix,
 superscriptaddress,
 nolongbibliography
]{revtex4-1}

\usepackage[T1]{fontenc}
\usepackage[utf8]{inputenc}
\usepackage{graphicx}
\usepackage{xcolor}
\usepackage{amsfonts,amsthm}
\usepackage{algorithm}
\makeatletter
\def\ftype@algorithm{8}
\@ifundefined{fbox@8}{\expandafter\newbox\csname fbox@8\endcsname}{}
\AtBeginDocument{\setcounter{float@type}{16}}
\makeatother
\usepackage{algpseudocode}
\usepackage{hyperref}
\hypersetup{colorlinks=true,allcolors=blue}

\newtheorem{Assertion}{Assertion}

\begin{document}

\title{Rate-selective synchronization in switching networks: when timing meets structure}

\author{Russell Jeter}
\affiliation{Department of Mathematics and Statistics, Georgia State University, Atlanta, GA 30303, USA.}

\author{Angelo Boateng}
\affiliation{Department of Mathematics and Statistics, Georgia State University, Atlanta, GA 30303, USA.}

\author{Maurizio Porfiri}
\affiliation{Center for Urban Science and Progress, Tandon School of Engineering, New York University, Brooklyn, New York 11201, USA.}
\affiliation{Department of Mechanical and Aerospace Engineering, Tandon School of Engineering, New York University, Brooklyn, New York 11201, USA.}
\affiliation{Department of Biomedical Engineering, Tandon School of Engineering, New York University, Brooklyn, New York 11201, USA.}

\author{Igor Belykh}
\email[Corresponding author: ]{ibelykh@gsu.edu}
\affiliation{Department of Mathematics and Statistics, Georgia State University, Atlanta, GA 30303, USA.}
\affiliation{Neuroscience Institute, Georgia State University, Atlanta, GA 30303, USA.}

\begin{abstract}
Switching networks arise across physics, biology, and engineering, where interactions turn on and off due to communication limits, control protocols, failures, or scheduling. Most synchronization theory for switching networks of continuous-time oscillators relies on the fast-switching approximation, yet many systems operate outside this regime, where switching can produce unexpected and counterintuitive collective dynamics. We develop the first analytical framework for synchronization in non-fast-switching networks of saddle-focus continuous-time oscillators. Using R\"ossler and piecewise-smooth systems as canonical examples, we leverage a structural property of saddle-focus oscillators to reduce the variational dynamics to a linear system with constant coefficients, enabling a Floquet-theoretic stability characterization under periodic switching.
The analysis identifies explicit switching intervals that stabilize synchronization even when neither switching mode nor the corresponding average system supports synchronization in isolation, thereby revealing the mechanism underlying the emergence of “windows of opportunity” at intermediate switching rates.
We further derive
a switching-dependent master stability function
linking the network Laplacian spectrum to the switching rate. The master stability function provides two complementary routes to
inducing or preserving synchronization: adjusting the switching
schedule for a fixed network or adapting the network spectrum to a
prescribed schedule through our constructive edge-pruning or edge-augmentation algorithms. These results provide a rigorous basis for understanding and leveraging intermediate-rate switching in the design and control of evolving networks.

\end{abstract}

\keywords{Floquet theory, master stability function, non-fast switching, oscillators, synchronization, temporal and blinking networks}

\date{\today}
\maketitle

\section{Introduction}
Networks of dynamical systems provide a unifying framework for modeling complex phenomena across physics, engineering, chemistry, biology, and the social sciences. \cite{strogatz2001exploring,newman2003structure} These models describe how local interactions among individual units can give rise to intricate, often nonlinear, collective dynamics. In recent years, substantial effort has been devoted to characterizing the algebraic, statistical, and graph-theoretical properties of networks and to understanding how these structural features influence system behavior. \cite{boccaletti2006complex,estrada2012structure} A key focus has been synchronization, which plays a critical role in the function and, in some cases, the dysfunction of a wide range of natural and technological systems.\cite{pikovsky2003synchronization,arenas2008synchronization}
In particular, significant attention has been devoted to the interplay between node dynamics and network topology, which fundamentally determines the stability and emergence of synchronization.
\cite{pecora1998master,boccaletti2002synchronization,barahona2002synchronization,wah2002synchronization,wang2002synchronization,belykh2004connection,belykh2006generalized,wu2006synchronization,li2006global,belykh2005synchronizationOfBursting,belykh2008weak,sun2009master,nishikawa2010network,aguirre2014synchronization,zhang2017incoherence,nishikawa2017sensitive,belykh2015synergistic,reimbayev2017two,daley2020synchronizability,honari2025spectral}

While much research has focused on networks with static topologies, where connections and coupling strengths remain fixed over time, many real-world systems are inherently dynamic. In such systems, the interaction topology may evolve in response to internal dynamics, external inputs, or environmental changes. Networks with time-varying connectivity, commonly referred to as temporal or evolving networks, have only recently gained sustained attention in the scientific literature. \cite{gorochowski2010evolving,delellis2010evolution,stojanovski1997sporadic,ito2001spontaneous,belykh2004blinking,hasler2005blinking,skufca2003communication,lu2005time,lu2007adaptive,mondal2008rapidly,chen2009enhanced,zanette2004dynamical,porfiri2006random,porfiri2008synchronization,porfiri2008master,porfiri2009global,porfiri2012stochastic,porfiri2011master,gorochowski2010evolving,delellis2010evolution,yu2012distributed,delellis2010synchronization,de2008synchronization,sorrentino2008adaptive,so2008synchronization,mucha2010community,hasler2013dynamicsPart1,hasler2013dynamicsPart2,belykh2013multistable,dorogovtsev2002evolution,gorochowski2012evolving,abaid2011consensus,frasca2008synchronization,delellis2013adaptive,masuda2013temporal} These studies have begun to uncover how time-dependent structures influence collective behavior, stability, and information flow in networks with dynamic architecture. For comprehensive treatments of this rapidly developing field, see the reviews. \cite{belykh2014evolving,holme2012temporal,holme2013temporal,masuda2017temporal,belykh2023beyond}

A particularly important class of evolving networks consists of systems with on-off, or switched, connectivity, in which links between nodes activate and deactivate over time. These switched networks represent a natural subclass of temporal networks and are prevalent in a variety of domains, including theoretical physics, neuroscience, and engineering. \cite{Liberzon2012} In biological systems, for example, discontinuous spatial rearrangements among components drive the temporal organization of interactions in active matter, leading to dynamically evolving connection patterns. \cite{camazine2003self,holme2012temporal,needleman2017active,feinerman2018physics} Similarly, in brain networks, synchronization arises through intermittent synaptic coupling among spiking and bursting neurons, with connections switching on and off in a temporally irregular fashion. \cite{churchland2016computational} Beyond natural systems, switch-based dynamics also underpin a wide range of engineered systems. Power converters, communication networks, and digital control systems often rely on switch logic to maintain stability and performance in time-varying operational environments. \cite{tse2002complex,o2001application}

Blinking networks represent an important class of dynamical systems and networks with on-off switching connections. \cite{belykh2004blinking,hasler2005blinking,porfiri2006random} In such networks, connections between nodes switch on and off much faster than the intrinsic timescale of the node dynamics. This fast-switching regime allows for analytical tractability and has been used extensively to study phenomena such as synchronization, consensus, and multistability in both continuous-time and discrete-time oscillator networks. \cite{belykh2014evolving} In this regime, the switching network's behavior closely approximates that of a static network with average, time-independent connections.
A mathematically rigorous theory for synchronization in fast-switching blinking networks has been developed in a number of studies. \cite{belykh2004blinking,hasler2005blinking,skufca2003communication,porfiri2006random,porfiri2008synchronization,porfiri2008master,porfiri2009global,porfiri2012stochastic,porfiri2011master,hasler2013dynamicsPart1,hasler2013dynamicsPart2,belykh2013multistable,jeter2015synchronization} These results provide important insights into how rapidly changing topologies influence the emergence and stability of collective dynamics. Beyond network synchronization, rigorous theory for the behavior of stochastically fast-switching networks based on Lyapunov functions and large deviation bounds for random variables has been developed in Refs.\cite{hasler2013dynamicsPart1,hasler2013dynamicsPart2,belykh2013multistable}. In particular, this theory yields explicit bounds connecting the probability of converging to an attractor in a multistable blinking network, the switching frequency, and the chosen initial conditions. \cite{hasler2013dynamicsPart2} However, extending the theory to dynamics and synchronization in networks with non-fast connections remains a significant challenge and is often out of reach. In such settings, the system dynamics do not follow the dynamics of the average system, and rigorous synchronization analysis becomes considerably more complex.

Non-fast-switching connectivity gives rise to a wide range of unexpected dynamical behaviors that the fast-switching limit does not capture. For instance, in multistable systems, certain stochastic switching sequences and specific optimal switching frequencies can cause the system’s trajectory to converge to a ghost attractor that is not invariant under the dynamics of the average system.  \cite{belykh2013multistable} Another striking phenomenon is the existence of bounded intervals of intermediate switching frequencies, referred to as ``windows of opportunity,'' within which synchronization becomes stable, even though it may be absent in both the fast-switching and slow-switching regimes. \cite{jeter2015synchronization} As a result, networks that fail to synchronize under fast switching may achieve synchronization at non-fast switching rates, only to lose it again as the frequency decreases further. These counterintuitive effects have been first observed numerically in networks of continuous-time Rössler and Duffing oscillators, as well as in ecological models such as the Rosenzweig–MacArthur food chain. \cite{jeter2015synchronization,jeter2015synchrony} Their emergence underscores the need for a rigorous theoretical framework to explain synchronization behavior beyond the fast-switching limit. Progress in this direction has been made using blinking networks of discrete-time systems (maps) with non-fast switching, \cite{golovneva2017windows,jeter2018Network1,jeter2018overcoming,porfiri2019windows} including those with Markovian switching schemes that incorporate memory effects. \cite{porfiri2017Memory} In such discrete-time settings, the switching period can be interpreted as the number of iterations of the individual map, allowing for a rescaled time representation. This formulation enables a mathematically rigorous framework for analyzing stochastic synchronization stability as a function of the switching period.

Despite growing numerical evidence, analytical understanding of windows of opportunity in non-fast-switching networks of continuous-time oscillators remains limited. This gap largely stems from the inherent complexity of switching dynamics in nonlinear continuous-time systems and the lack of suitable stability methods.

Notable exceptions are Refs. \cite{porfiri2021antiresonance, xiedeng2026dynamic}, which present analytically tractable, counterintuitive examples of non-fast switched systems that alternate between two unstable modes yet remain stable over a finite range of switching frequencies. In Ref. \cite{porfiri2021antiresonance}, a two-dimensional switched continuous-time system was shown to have unstable average dynamics, excluding stability in the fast-switching regime, while both individual modes are also unstable, excluding stability at low switching rates. Despite this, the system stabilizes over an intermediate frequency range, forming a window of opportunity through the interplay of unstable dynamics. The mechanism is governed by a fixed point that simultaneously exhibits characteristics of an unstable node and a stable focus, a phenomenon termed anti-resonance stabilization. Likewise, Ref.\cite{xiedeng2026dynamic} introduced a mechanical oscillator that switches between two unstable modes, has an unstable average, and yet can still be dynamically stabilized. Based on a modified Meissner model, the system modulates both stiffness and damping through square-wave switching between positive and negative values. The work theoretically established compact anti-resonance windows and experimentally validated them using a cantilever-beam oscillator under magnetic and aerodynamic forcing. Although these studies rigorously characterize anti-resonance phenomena in switched systems with fixed-point stabilization, comparable analytical results for synchronization in continuous-time oscillator networks are still lacking. In particular, proving the existence of windows of opportunity in such networks and rigorously characterizing how synchronization depends on non-fast switching rates have remained elusive.

In this work, we address this long-standing challenge by providing the first analytical treatment of windows of opportunity for synchronization in non-fast switching networks of saddle-focus oscillators.  These oscillators exhibit Shilnikov-type chaos,\cite{shilnikov1965case} organized around a saddle-focus equilibrium with a two-dimensional unstable focal manifold and a one-dimensional stable manifold. We focus on networks of R\"ossler oscillators,\cite{rossler1976equation} a canonical model in synchronization theory and the original system used to introduce foundational stability tools such as the master stability function. \cite{pecora1998master} Our approach builds on recent advances \cite{tang2022multilayer} showing that the transverse variational dynamics of statically coupled oscillators in this class
can be approximated by a constant-coefficient linear system describing an unstable two-dimensional focus. Stabilizing this focal subsystem is sufficient
to ensure synchronization stability, a structural property shared by
R\"ossler oscillators and certain piecewise-smooth systems.
\cite{barajas2021generating} This reduction transforms the original
time-varying variational problem, which generally requires numerical
Lyapunov exponent computations, into an analytically tractable
stability analysis with explicitly computable eigenvalues.

Building on this framework, we show that synchronization in non-fast switching networks can be analyzed analytically. In the most representative parameter regimes, the reduced linear system alternates between an unstable focus and a saddle. By applying Floquet theory, \cite{rugh1996linear} we construct the transition matrix over one switching period and characterize its eigenvalues. This analysis identifies explicit switching intervals that stabilize the system, and synchronization emerges in the original on-off switching network. Extending the analysis to larger networks, we uncover a nontrivial interplay between network topology and non-fast switching periods that support synchronization, an effect previously accessible only through numerical simulation. Moreover, we derive a constructive analytical algorithm that adapts the network topology via edge removal to induce or preserve stable synchronization as the switching rate changes. Finally, we prove that the same mechanism extends beyond Rössler dynamics to networks of piecewise-smooth saddle-focus oscillators, \cite{barajas2021generating} which exhibit synchronized chaotic behavior under non-fast switching.

\section{Switching Network Model}
We consider a switching network of $N$ identical oscillators,
   \begin{equation}
    \frac{d \mathbf{x}_i}{dt}= \mathbf{F}(\mathbf{x}_i) + d \sum\limits_{j = 1}^N a_{ij}^{(s(t))} P(\mathbf{x}_j-\mathbf{x}_i), \quad i=1,\ldots,N,
    \label{model}
    \end{equation}
where $\mathbf{x}_i\in \mathbb{R}^m$ is the state of oscillator $i$, $\mathbf{F}$ governs the intrinsic oscillator dynamics, $d>0$ is the uniform coupling strength, and $P$ is an $m\times m$ projection matrix selecting the variables through which the oscillators are coupled. The matrix $A^{s(t)}=\{a_{ij}^{s(t)}\}$ describes the active network configuration at time $t$, and the corresponding graph Laplacian is denoted by $L^{s(t)}$. {The switching signal $s(t)$ selects one of two connectivity states, $s_1$ or $s_2$, with corresponding Laplacians $L^{s_1}$ and $L^{s_2}$. These labels denote arbitrary network configurations; both Laplacians may be nonzero.}
We focus on switching between two matrices for clarity; the same formulation extends to periodic switching sequences involving three or more network configurations. One switching mode may also be disconnected, including the limiting
case $L^{s_1}=\mathbf{0}_{N\times N}$, corresponding to an edgeless
network with all coupling switched off during that phase.

\begin{figure}[!tbp]
	\centering
	\includegraphics[width=\linewidth]{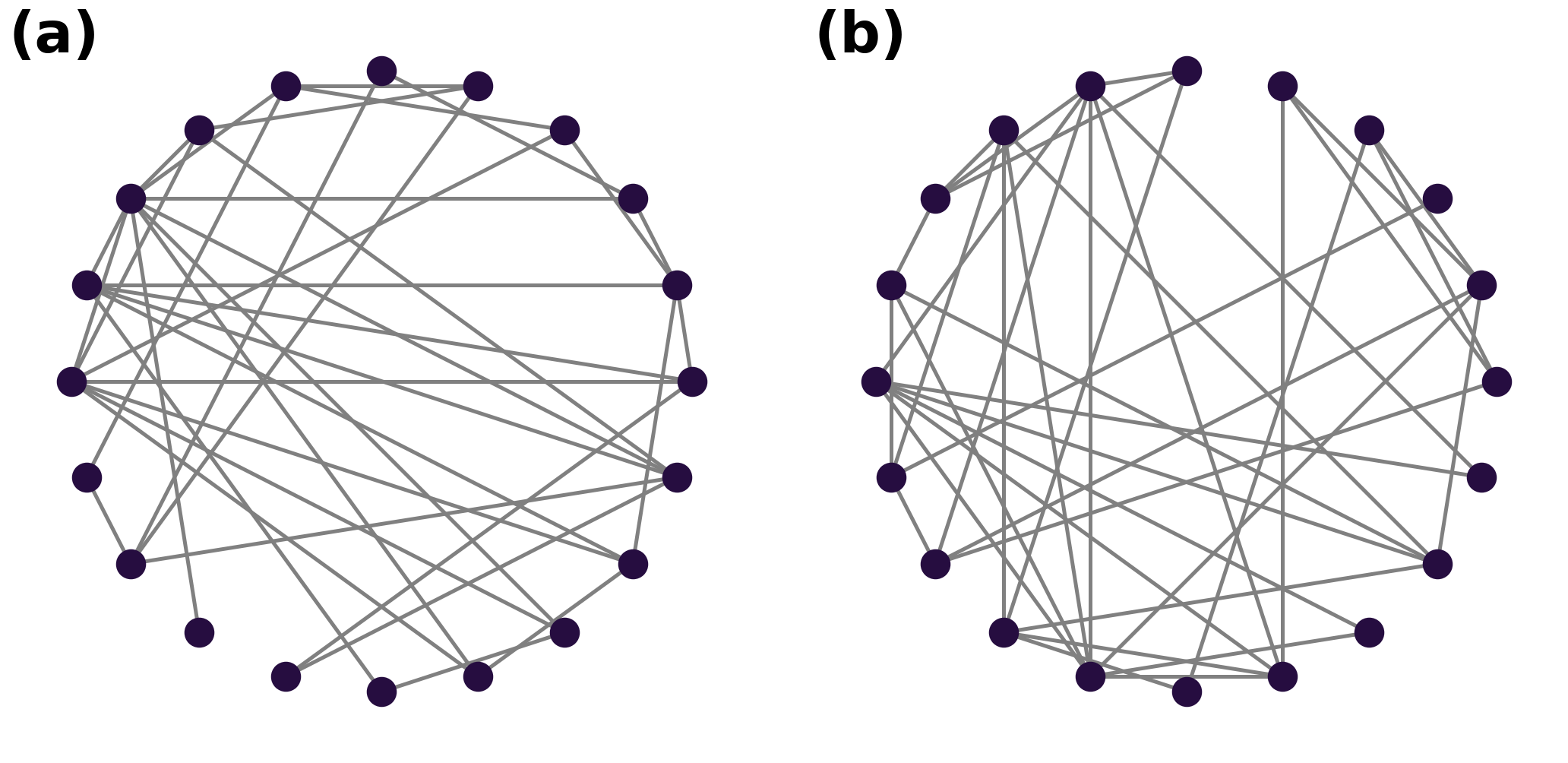}
	\caption{A $20$-node switching network alternating periodically between two connectivity configurations, each an independently generated, connected Erd\H{o}s--R\'{e}nyi random graph with edge probability $p=0.2$. (a) Mode $s_1$, active for $\delta T$ units of time. (b) Mode $s_2$, active for the remaining $(1-\delta)T$, after which the cycle repeats.}
  \label{20 node on-off}
\end{figure}

For the analytical development below, we assume $s(t)$ is periodic with period $T$ and duty cycle $\delta$: the network is in mode $s_1$ for $\delta T$ and in mode $s_2$ for $(1-\delta)T$. Figure~\ref{20 node on-off} illustrates switching between the two connectivity matrices for a network of $20$ oscillators.

Of particular interest is the relationship between the dynamics of the switching network \eqref{model} and the dynamics of its average counterpart. The average system is the static network obtained by replacing the switching connectivity with its period-average,
\begin{equation}
    {L^{\rm avg}=\delta L^{s_1}+(1-\delta)L^{s_2}.}
  \label{static}
\end{equation}
In the fast-switching regime, where $T$ is much smaller than the characteristic time scale of the individual oscillators, solutions of the switching network are well approximated by solutions of the average system, enabling analytical characterization of collective behavior.  \cite{belykh2004blinking,hasler2013dynamicsPart2} By contrast, when switching occurs on time scales comparable to the intrinsic dynamics, the average description may fail. This non-fast regime is the focus of the present work, aiming at an analytical understanding of the emergence of windows of opportunity for stabilizing the strongest form of collective dynamics in networks  \eqref{model}, complete synchronization defined by
the synchronization manifold
$\mathcal{M} = \left \{\mathbf{x}_i(t)=\mathbf{x}_j(t)=\mathbf{x}(t) \;\mbox{\rm for}\;\forall i,j \in\left\{1, 2, \ldots, N\right\} \right\}$.
This manifold is invariant for {both frozen switching modes $s_1$ and $s_2$} and for the average system with static connection graph \eqref{static}.
However, the switching rate controls its transversal stability in the switching network, and thus the stability of network synchronization, and it can exhibit counterintuitive phenomena. Specifically, synchronization may be unstable in each switching mode, and its average counterpart may also be unstable; yet, non-fast switching can robustly stabilize network synchronization over a fairly wide range of switching periods.
\section{Two-Node Switching Network}\label{sec:two_node}
To showcase rate-selective synchronization and its nontrivial stabilization mechanism in non-fast switching networks \eqref{model}, we introduce a minimal, two-node switching network of $x$-coupled chaotic R\"ossler oscillators. \cite{pecora1998master} The network alternates between an ``on'' configuration with active coupling and an ``off'' configuration in which the oscillators are fully decoupled.
The switching network is described by
\begin{equation}
\begin{array}{l}
    \dot{x}_1 = -y_1 - z_1 + d s(t) \left(x_2 - x_1 \right),\\
    \dot{y}_1 = x_1 + ay_1,\\
    \dot{z}_1 = b + \left(x_1 - c\right)z_1,\\
    \dot{x}_2 = -y_2 - z_2 + d s(t) \left(x_1 - x_2\right),\\
    \dot{y}_2 = x_2 + ay_2,\\
    \dot{z}_2 = b + \left(x_2 - c\right)z_2,
    \end{array}
    \label{two-rossler}
\end{equation}
where $s(t)$ represents periodic on-off switching: $s(t)=0$ for $\delta T$ units of time and $s(t)=1$ for $(1-\delta)T$ units of time during each period $T$.
The parameters $a=b=0.1$ and $c=14.0$ place the uncoupled R\"ossler oscillator in a chaotic saddle-focus regime. \cite{rossler1976equation,shilnikov1965case} The parameter $d$ is the strength of the diffusive coupling between the $x$ variables.

The corresponding average system is obtained from \eqref{two-rossler} by replacing the switching variable $s(t)$ with its mean $1-\delta$:
\begin{equation}
\begin{array}{l}
    \dot{x}_1 = -y_1 - z_1 + d_{\rm avg} \left(x_2 - x_1 \right),\\
    \dot{y}_1 = x_1 + ay_1,\\
    \dot{z}_1 = b + \left(x_1 - c\right)z_1,\\
    \dot{x}_2 = -y_2 - z_2 + d_{\rm avg}\left(x_1 - x_2\right),\\
    \dot{y}_2 = x_2 + ay_2,\\
    \dot{z}_2 = b + \left(x_2 - c\right)z_2,
    \end{array}
    \label{average}
\end{equation}
where $d_{\rm avg}=(1-\delta)d.$

In the regime of complete synchronization, defined by
the synchronization manifold
$\mathcal{M} = \{x_1=x_2=x(t),\; y_1=y_2=y(t),\; z_1=z_2=z(t)\}$,
  the synchronous dynamics on the manifold when all diffusive coupling terms vanish is defined by  a single uncoupled R\"ossler oscillator,
\begin{equation}
    \dot{x}=-y-z,\qquad \dot{y}=x+ay,\qquad \dot{z}=b+(x-c)z.    \label{synchronous_solution_rossler}
\end{equation}

Before analyzing the switching network \eqref{two-rossler}, we first examine the transverse stability of its average counterpart \eqref{average}. This static problem provides the natural reference point for the fast-switching limit and the baseline against which we compare the non-fast-switching effect.

\subsection{Transverse variational dynamics of the average system}
\label{subsec:averaged_variational_dynamics}
Following the standard difference-variable approach, we introduce the infinitesimal transverse perturbations
\begin{equation}
    \xi=x_2-x_1,\qquad \eta=y_2-y_1,\qquad \zeta=z_2-z_1 .
    \label{two_node_difference_variables}
\end{equation}
Linearizing the average system \eqref{average} about the synchronization manifold \(\mathcal{M}\), with the derivatives evaluated along the synchronous solution \((x(t),y(t),z(t))\) of \eqref{synchronous_solution_rossler}, yields the transverse variational equations
\begin{equation}
    \begin{aligned}
    \dot{\xi} &= -\eta-\zeta-2 d_{\rm avg}\xi,\\
    \dot{\eta} &= \xi+a\eta,\\
    \dot{\zeta} &= z(t)\xi+[x(t)-c]\zeta .
    \end{aligned}
    \label{avg_full_variational_system}
\end{equation}

\begin{figure}[!tbp]
    \centering
    \includegraphics[width=0.80\linewidth]{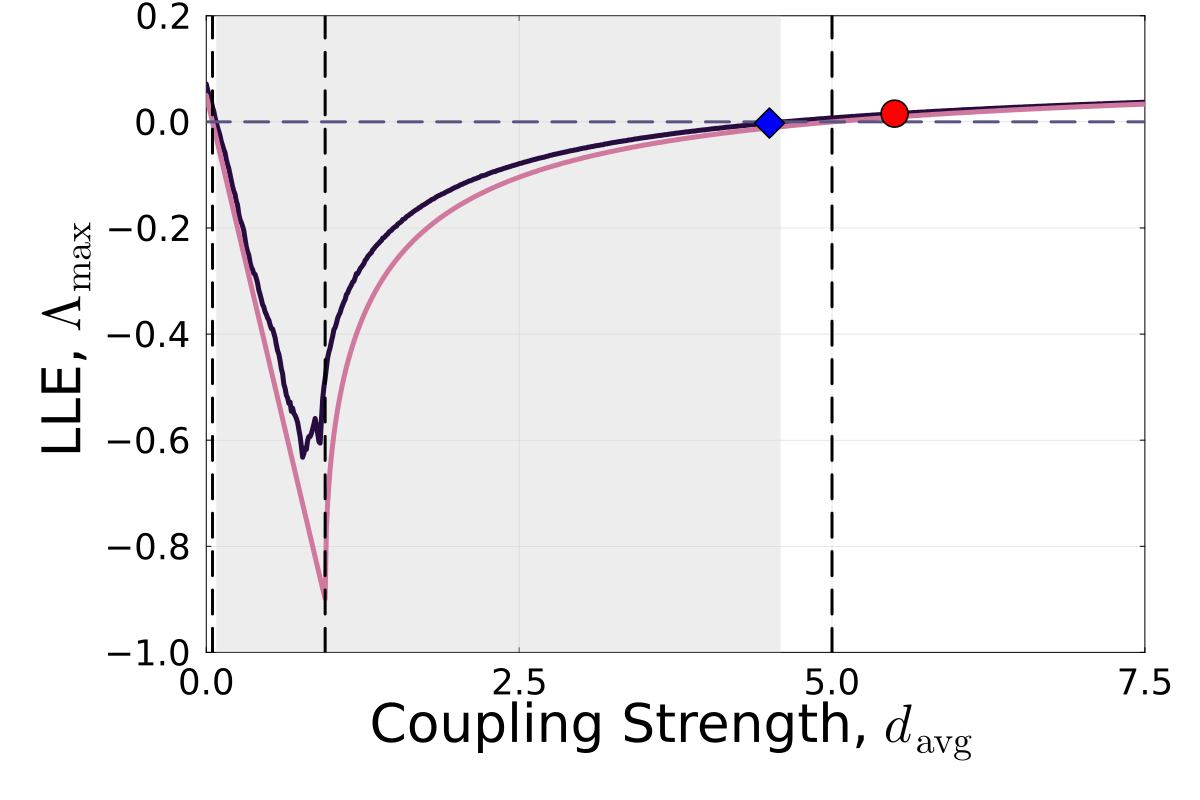}
        \caption{Synchronization stability in the average two-node system \eqref{average}. The purple curve shows the largest transverse Lyapunov exponent computed from the full variational equations \eqref{avg_full_variational_system}, and the pink curve shows the largest real part of the eigenvalues of the reduced average system \eqref{avg_reduced_variational_system}. Negative values indicate stable synchronization. The bounded interval of negative values is the average system's synchronization window. For \(\delta=1/2\), the switching strength \(d=11\) used in Fig.~\ref{Switching Period Comparison} gives \(d_{\rm avg}=5.5\), marked by the red circle; this point lies beyond the upper boundary of the static synchronization window, where coupling is too strong. The blue diamond marks the stable average coupling used for comparison in Fig.~\ref{ILE color coded attractors}.}
    \label{LLE Numerical vs. Analytical}
\end{figure}

The variational system \eqref{avg_full_variational_system} has time-dependent coefficients through the synchronous R\"ossler attractor trajectory. Therefore, the stability of the synchronization manifold $\mathcal{M}$ in the average system is assessed by the largest  Lyapunov exponent (LLE), \(\Lambda_{\max}\) of \eqref{avg_full_variational_system}.
Networks of $x$-coupled R\"ossler oscillators are known to exhibit stable synchronization only in a bounded interval of coupling strength, and synchronization is lost when the coupling becomes either too weak or too strong. \cite{pecora1998master} Figure~\ref{LLE Numerical vs. Analytical} shows this bounded stability window for the average system \eqref{average}.

To obtain an analytically tractable approximation of the variational system \eqref{avg_full_variational_system}
that further enables analysis of the switching system, we use the saddle-focus reduction introduced in Ref.~\cite{tang2022multilayer}. The key observation is that the transverse stability is largely determined by the focal component of the synchronous motion. Near the focal part of the attractor, the dynamics is effectively confined to the \((x,y)\)-plane, where the R\"ossler equations reduce to the linear planar flow
\begin{equation}
    \dot{x}=-y,\qquad \dot{y}=x+a y,
    \label{rossler_focal_plane_flow}
\end{equation}
whereas the \(z\)-direction gives the one-dimensional stable component of the saddle-focus. Thus, the dominant contribution to transverse contraction or expansion can be captured by the \((\xi,\eta)\) perturbations associated with this focal planar dynamics.

\begin{figure}[!tbp]
    \centering
    \includegraphics[width=0.95\linewidth]{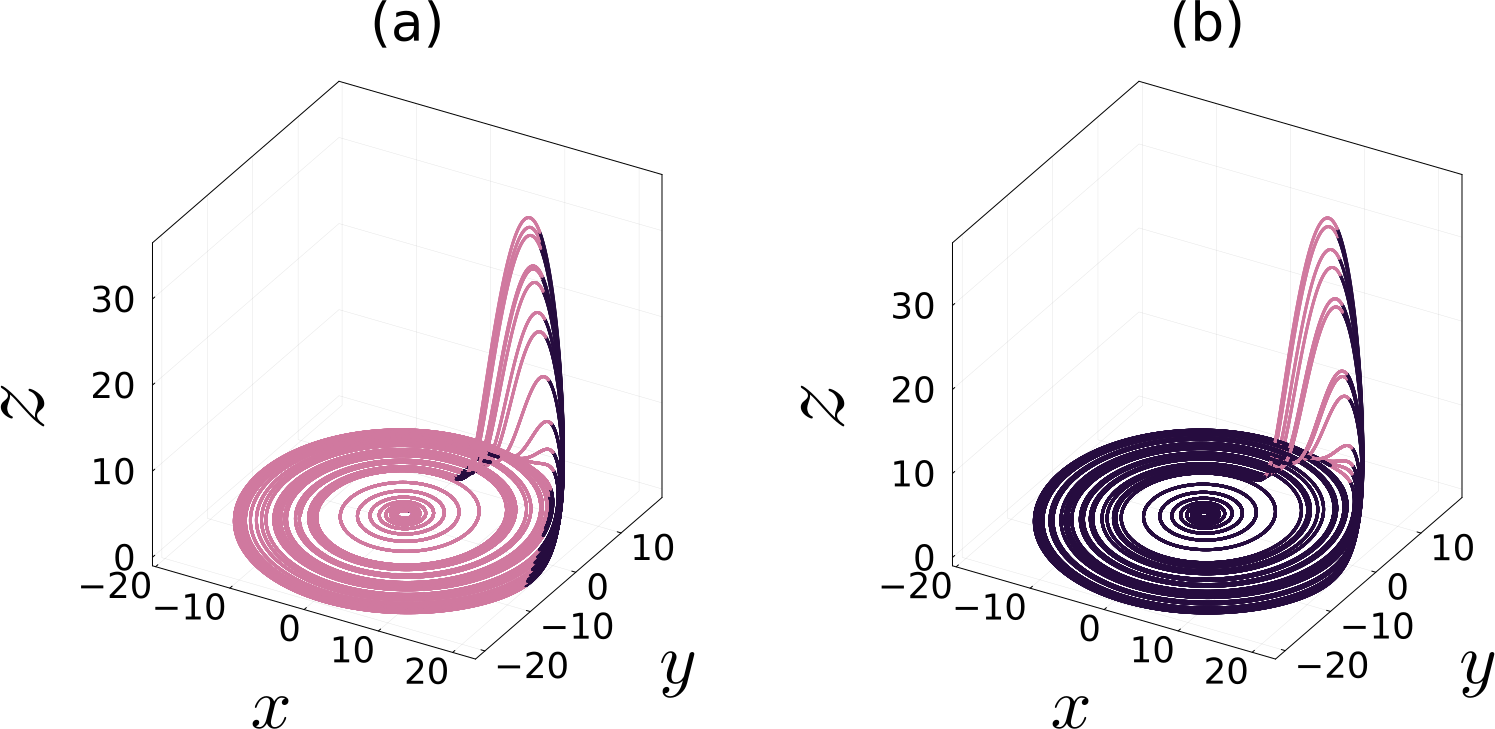}
        \caption{Instantaneous convergence and divergence of the variational system \eqref{avg_full_variational_system} along the synchronous R\"ossler trajectory of the average two-node system \eqref{average} for (a) \(d_{\rm avg} = 4.5\) and (b) \(d_{\rm avg} = 5.5\). These coupling strengths are marked by the blue diamond and red circle, respectively, in Fig.~\ref{LLE Numerical vs. Analytical}. Colors represent the binarized sign of the instantaneous LLE, computed from the variational system using the same procedure as in Fig.~\ref{LLE Numerical vs. Analytical}. Dark purple denotes positive values (instantaneous divergence from the synchronous solution), while pink denotes negative values (instantaneous convergence toward it).}
  \label{ILE color coded attractors}
\end{figure}

Figure~\ref{ILE color coded attractors} illustrates this state-dependent balance between instantaneous convergence and divergence for the average two-node system. Figure~\ref{ILE color coded attractors}(a) corresponds to \(d_{\rm avg}=4.5\), a coupling value inside the synchronization window of Fig.~\ref{LLE Numerical vs. Analytical}; Figure~\ref{ILE color coded attractors}(b) corresponds to \(d_{\rm avg}=5.5\), beyond the upper boundary of that window. Pink portions of the attractor mark instantaneous convergence toward the synchronization manifold, while dark purple portions mark instantaneous divergence away from it. The comparison shows that changing the coupling strength changes how convergence and divergence are distributed along the same saddle-focus geometry. For \(d_{\rm avg}=4.5\), the contracting portions dominate the net transverse growth and the average system synchronizes. For \(d_{\rm avg}=5.5\), the balance shifts toward net expansion, and synchronization is lost. The close alignment of these convergence and divergence regions with the focal part of the trajectory supports using the reduced \((\xi,\eta)\) dynamics to predict the full variational system~\eqref{avg_full_variational_system}.

Therefore, following Ref. \cite{tang2022multilayer}, we neglect the \(\zeta\)-perturbation associated with excursions away from the focal plane and reduce \eqref{avg_full_variational_system} to the constant-coefficient linear system
\begin{equation}
    \begin{aligned}
         \dot{\xi} &= -\eta - 2d_{\rm avg}\xi,\\
         \dot{\eta} &= \xi + a\eta.
    \end{aligned}
    \label{avg_reduced_variational_system}
\end{equation}
The pink curve in Fig.~\ref{LLE Numerical vs. Analytical} is the largest real part of the eigenvalues of \eqref{avg_reduced_variational_system}. Its close agreement with the full Lyapunov exponent curve shows that the reduced average system predicts the stability boundaries of the full R\"ossler variational dynamics with high accuracy. Thus, the reduced system retains the essential bounded synchronization window of the static average network. The next question is whether the same reduced description can explain why a switching network may synchronize even when the corresponding average system is unstable.

\subsection{A synchronization puzzle beyond fast switching}
The central question is what happens when the static connection is replaced by a periodic on-off connection that alternates between the disconnected mode $s(t)=0$ and the connected mode $s(t)=1$. We track the resulting dynamics through the synchronization error
\begin{equation}
    \Delta_{\rm sync}(t)=\frac{2}{mN(N-1)}\sum_{i<j}^{N}\bigl\|\mathbf{x}_i(t)-\mathbf{x}_j(t)\bigr\|_1 ,
    \label{eq:network_sync_error}
\end{equation}
which measures the mean absolute difference between corresponding state components. For the present two-node network of three-dimensional R\"ossler oscillators, this reduces to $\Delta_{\rm sync}=\bigl(|x_1-x_2|+|y_1-y_2|+|z_1-z_2|\bigr)/3$.

\begin{figure}[!tbp]
	\centering
	\includegraphics[width=0.90\linewidth]{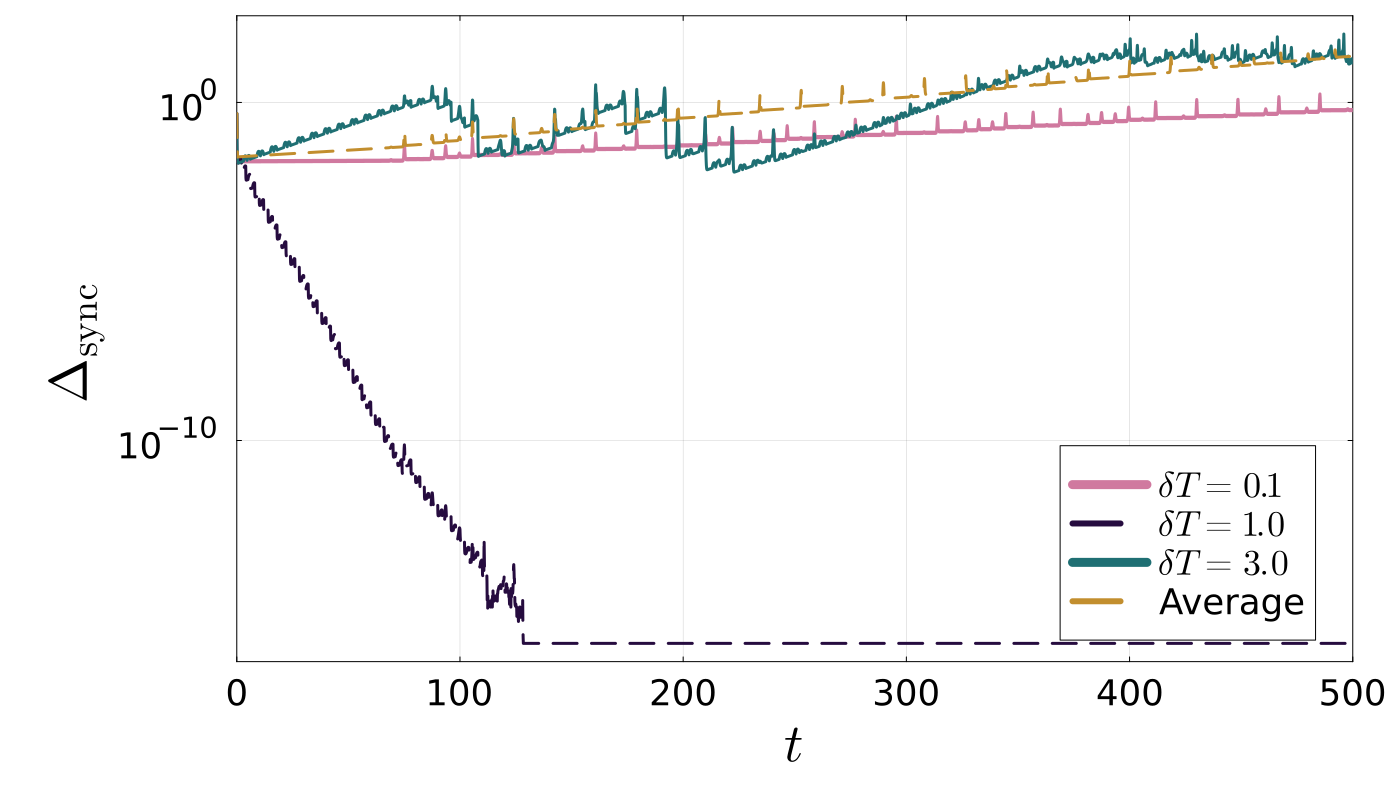}
		\caption{Synchronization error $\Delta_{\rm sync}(t)$, defined in \eqref{eq:network_sync_error}, for two periodically switched, $x$-coupled R\"ossler oscillators in \eqref{two-rossler}, with $d=11$ and $\delta=1/2$. The average system \eqref{average} has effective coupling $d_{\rm avg}=5.5$, corresponding to the red circle in Fig.~\ref{LLE Numerical vs. Analytical}, where the static/average system is unstable because the coupling is too strong. Accordingly, fast switching ({$\delta T=0.1$}) fails to synchronize. Slow switching ({$\delta T=3.0$}) is also unstable because the system spends long intervals in the disconnected mode, during which the two oscillators evolve independently. In contrast, the non-fast {switching interval $\delta T=1.0$} stabilizes synchronization, revealing a window of opportunity. {The dashed gold curve corresponds to the average system \eqref{average}.}}
  \label{Switching Period Comparison}
\end{figure}

Figure~\ref{Switching Period Comparison} demonstrates the puzzle that motivates the rest of the paper. In the fast-switching regime, the switching system closely follows the average system \eqref{average}; hence the stability condition is inherited from the static curve in Fig.~\ref{LLE Numerical vs. Analytical}. With $d=11$ and $\delta=1/2$, this average coupling is $d_{\rm avg}=5.5$, the red circle in Fig.~\ref{LLE Numerical vs. Analytical}, where the coupling is already beyond the upper edge of the bounded synchronization interval. Synchronization in the fast-switching network is therefore unstable. In the slow-switching regime, the disconnected phase persists long enough for the two oscillators to drift apart, and synchronization is again lost. Nevertheless, for the non-fast {switching interval $\delta T=1.0$}, the switching sequence stabilizes the synchronous manifold even though neither the fast-switching limit nor the slow-switching limit supports synchronization.  Next, we reveal the mechanism by which such windows emerge from non-fast switching.

\subsection{Dynamical stabilization mechanism under non-fast switching}\label{geometric}

\begin{figure}[!tbp]
	\centering
	\includegraphics[width=\linewidth]{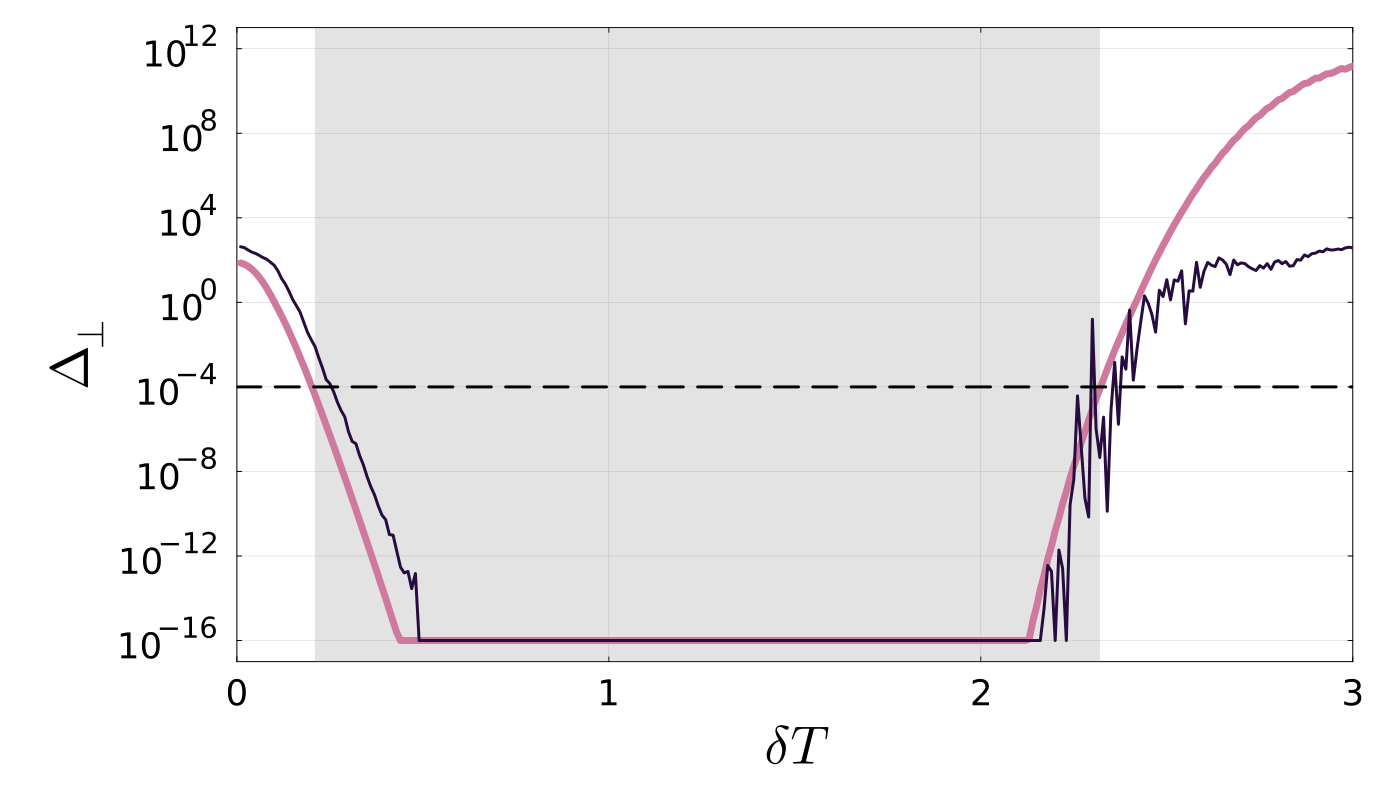}
		\caption{Windows of opportunity for the full three-dimensional variational system \eqref{switching_full_variational_system} (dark purple) and its two-dimensional reduction \eqref{reduced_s} (pink) under periodic on--off switching between $s_1=0$ and $s_2=1$, and with $d=11$ and $\delta=1/2$. {The numerical curves show the transverse error $\Delta_{\perp}$, the maximum of the summed magnitudes of all transverse perturbations, $|\xi(t)|+|\eta(t)|+|\zeta(t)|$ for the full system and $|\xi(t)|+|\eta(t)|$ for the reduced system,} measured over the final 50 time units, as a function of the switching interval $\delta T$ for duty cycle $\delta=1/2$. The shaded region marks a window of opportunity in which the reduced system is stable. The close agreement between the two curves demonstrates that the reduced system accurately predicts the corresponding stability window of the full variational dynamics. Additional windows recur at longer switching intervals. This continuous parameter sweep complements Fig.~\ref{Switching Period Comparison}, which illustrates the same phenomenon at three representative switching intervals.}
	\label{Reduced Rossler window sweep}
\end{figure}

As in the average-system analysis, we linearize the switching system
\eqref{two-rossler} about the synchronous solution. The transverse
perturbations $\xi$, $\eta$, and $\zeta$, defined in
\eqref{two_node_difference_variables}, satisfy
\begin{equation}
    \begin{aligned}
        \dot{\xi}  &= -\eta-\zeta-2d s(t)\xi,\\
        \dot{\eta} &= \xi+a\eta,\\
        \dot{\zeta}&=z(t)\xi+[x(t)-c]\zeta .
    \end{aligned}
    \label{switching_full_variational_system}
\end{equation}
Following the same focal-plane approximation used for the average
system, we neglect the $\zeta$ dynamics and its contribution to
$\dot{\xi}$. This yields the reduced switching system
\begin{equation}
    \begin{aligned}
        \dot{\xi}  &= -\eta-2d s(t)\xi,\\
        \dot{\eta} &= \xi+a\eta.
    \end{aligned}
    \label{reduced_s}
\end{equation}
Figure~\ref{Reduced Rossler window sweep} compares the transverse
stability of the full variational system
\eqref{switching_full_variational_system}
and its reduced counterpart
\eqref{reduced_s} for on--off switching between $s_1=0$ and $s_2=1$. It
demonstrates the existence of a finite interval of stabilizing switching periods (gray regions), yielding a window of opportunity for stable synchronization.
\cite{jeter2015synchronization,jeter2015synchrony} The close agreement between their stability boundaries throughout the
window of opportunity indicates that the reduced system captures the
essential mechanism responsible for stabilization. It can therefore be
used as an effective analytical model for explaining the emergence of
these windows.

\begin{figure}[!tbp]
	\centering
	\includegraphics[width=\linewidth]{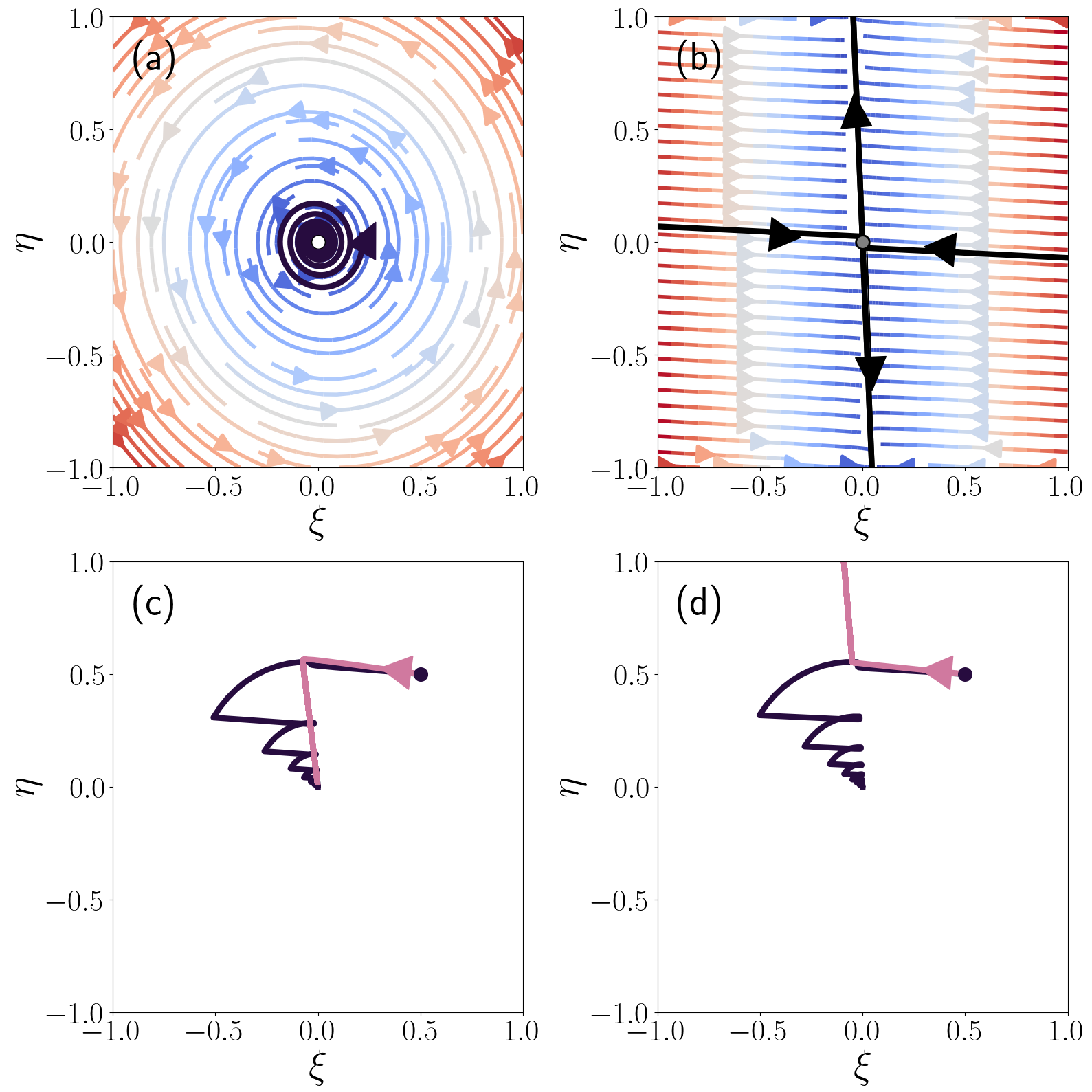}
		\caption{(a), (b) Phase portraits of the reduced variational system  \eqref{reduced_s} in the two regimes between which the switching system alternates. (a) The uncoupled mode $d=0$ is an unstable focus at the origin; the dark purple curve is a trajectory spiraling away from it. (b) The coupled mode $d=11$, that is $-d\gamma_l = 22$, is a saddle at the origin; the black lines indicate its stable and unstable directions. Streamlines are colored by the magnitude of the vector field. (c), (d) Trajectories of the on-off switched reduced system started from $(\xi_0,\eta_0)=(0.5,0.5)$ for fast switching $\delta T = 0.001$ (pink) and non-fast switching $\delta T = 1$ (dark purple), with duty cycle $\delta = 1/2$. (c) $d = 7.5$, so that $d_{\rm avg} = 3.75$ lies inside the synchronization window of Fig.~\ref{LLE Numerical vs. Analytical} and both trajectories converge to the origin. (d) $d = 11$, so that $d_{\rm avg} = 5.5$ lies beyond the upper edge of that window (the red circle in Fig.~\ref{LLE Numerical vs. Analytical}): the fast-switching trajectory diverges, as the average system predicts, whereas the non-fast-switching trajectory still converges to the origin. Panel (d) is the window of opportunity of Fig.~\ref{Switching Period Comparison} reproduced in the reduced system. The average system is not drawn separately in (c) and (d) because its trajectory is indistinguishable from the fast-switching one at this scale.}
  \label{Periodic switching trajectories for the variational system}
\end{figure}

Let $\mathbf{u}=(\xi,\eta)^{\mathsf T}$. The reduced system
\eqref{reduced_s} alternates between two linear modes,
$\dot{\mathbf{u}}=A_{s(t)}\mathbf{u}$, governed by
\begin{equation}
    A_0=
    \begin{bmatrix}
        0 & -1\\
        1 & a
    \end{bmatrix},
    \qquad
    A_1=
    \begin{bmatrix}
        -2d & -1\\
        1 & a
    \end{bmatrix},
    \label{A0_A1}
\end{equation}
where $A_0$ corresponds to the off-mode, $s_1=0$, and $A_1$
corresponds to the on-mode, $s_2=1$. Their eigenvalues are
\begin{equation}
\begin{array}{l}
    \lambda_{1,2}^{A_0}
    =\frac{1}{2}\left(a\pm\sqrt{-4+a^2}\right),
    \\
    \lambda_{1,2}^{A_1}
    =\frac{1}{2}\left(
        a-2d\pm\sqrt{-4+a^2+4ad+4d^2}
    \right).
    \end{array}
\end{equation}

For the standard R\"ossler system's parameter $a=0.1$, the off-mode governed
by $A_0$ is an unstable focus (Fig.~\ref{Periodic switching trajectories for the variational system}(a)). The type of the on-mode governed by
$A_1$ depends on the coupling strength: it is an unstable focus for
$d<a/2$, a stable focus for
$a/2<d<(2-a)/2$, a stable node for
$(2-a)/2<d<1/(2a)$, and a saddle for $d>1/(2a)$.
Consequently, the value $d=11$ used to calculate the windows of
opportunity in Fig.~\ref{Reduced Rossler window sweep} places the on mode in the saddle
regime (Fig.~\ref{Periodic switching trajectories for the variational system}(b)). The variational dynamics therefore alternates between an
unstable focus and a saddle. During the off phase, a generic
perturbation spirals away from the origin. During the on phase, the
saddle contracts perturbations along its stable eigendirection but
expands them along its unstable eigendirection, so a generic
trajectory also eventually diverges if this mode remains active for
long enough; see Fig.~\ref{Periodic switching trajectories for the variational system}(a)--(b).

The two limiting switching regimes clarify why stabilization is
specific to an intermediate time scale. Under sufficiently fast
switching, the switching system closely follows its average
counterpart. When the coupling is chosen so that the average system
is stable, fast switching
therefore produces the expected stable behavior (Fig.~\ref{Periodic switching trajectories for the variational system}(c)). In contrast, the red circle in Fig.~2
corresponds to $d_{\rm avg}=5.5$, for which the average variational
system \eqref{avg_reduced_variational_system} is unstable. In the
reduced description, its coefficient matrix is a weak saddle, and its
transverse perturbations diverge. The fast-switching system inherits
this instability because it remains close to the average dynamics.
At the other extreme, when switching is slow, the system spends
sufficient time in each individual unstable mode for its instability
to dominate: perturbations spiral outward in the off mode or
 escape along the unstable eigendirection of the saddle in
the on mode.

Stabilization at intermediate switching periods results from the
directional contraction of the saddle combined with the rotational
dynamics of the unstable focus. Suppose that, during the off phase,
a perturbation follows an outward spiral arc generated by $A_0$. If
the switch occurs when this rotation has placed the perturbation close
to the stable eigendirection of $A_1$, the subsequent saddle dynamics
moves it toward the origin. The on phase must then end before the
component along the saddle's unstable eigendirection becomes dominant.
When the system switches back to the unstable-focus mode, the
perturbation begins another outward spiral, but now from a smaller
distance from the origin. If the next switch again occurs at a
favorable orientation, the perturbation is returned to the contracting
direction of the saddle. Repetition of this process produces a net
contraction over each complete switching cycle, and the trajectory
eventually converges to the origin even though neither mode is
asymptotically stable on its own.

The resulting window of opportunity is therefore controlled by the
interplay among the expansion rate of the unstable focus, the
contraction rate along the stable eigendirection of the saddle, the
growth rate along its unstable eigendirection, and the two dwell
times. The switching instants determine how the rotating perturbations are
oriented relative to the saddle's contracting direction and hence
whether successive phases produce net decay. Favorable phase alignment
can recur when the dwell time in the rotating phase is approximately
commensurate with the focal rotation period, for example, near one or
two full periods or a suitable fraction of a period. The precise
location and width of the resulting stability windows also depend on
the balance of expansion and contraction over the complete switching
cycle.

This geometric explanation connects the windows of opportunity to
the saddle-focus structure of the synchronous R\"ossler dynamics. It is analogous to the stabilization mechanism established for the switched mechanical oscillator in Ref.\cite{xiedeng2026dynamic},
where switching occurs between a saddle and an unstable node, with the node's unstable directions suitably modified to produce rotation-like dynamics.
To quantify the effect and predict the stability boundaries, we next
apply Floquet theory, construct the corresponding two-dimensional
monodromy matrix, and analytically compute its Floquet multipliers and
associated characteristic exponents.
\section{Master Stability Function for Switching Networks}
\label{sec:switching_msf}
We now turn the geometric explanation of Sec.~\ref{geometric} into analytical stability conditions for the reduced variational system. To avoid repeating the two-node network calculation, we formulate the analysis for an arbitrary undirected network topology, recovering the two-node example via its single nonzero Laplacian eigenvalue, $\gamma_2=-2$. Complete synchronization corresponds to the origin of the transverse variational dynamics. All transverse perturbations must decay for synchronization to be linearly stable. Our conditions are exact for the focal-plane reduction introduced in Sec.~\ref{sec:two_node}; their predictions for the original R\"ossler network are assessed by comparison with the full variational dynamics.

\subsection{Laplacian modes and reduced variational dynamics}
\begingroup
Each Laplacian $L^{s_1}$ or $L^{s_2}$ is symmetric and negative semidefinite. For any one fixed topology $L$, we order its eigenvalues as
\begin{equation}
    0=\gamma_1\geq\gamma_2\geq\cdots\geq\gamma_N.
    \label{eq:laplacian_spectrum}
\end{equation}
The eigenvector $\mathbf{1}_N/\sqrt{N}$ associated with $\gamma_1=0$ describes perturbations along the synchronization manifold; the remaining $N-1$ orthogonal modes are transverse to it. For a connected graph, $\gamma_2<0$, so all transverse eigenvalues are strictly negative. In the on--off special case, we set $s_1=0$ and $s_2=1$, with $L^{s_1}=L^{(0)}=\mathbf{0}_{N\times N}$ and $L^{s_2}=L$. The notation $L^{(0)}$ is reserved for this uncoupled mode. The same eigenbasis then applies throughout the switching cycle, irrespective of the topology of $L$.
\par
\endgroup

\begingroup
The calculation also applies to switching between two nonzero Laplacians $L^{s_1}$ and $L^{s_2}$, provided they share an orthonormal eigenbasis. For symmetric Laplacians, this is equivalent to $L^{s_1}L^{s_2}=L^{s_2}L^{s_1}$. We denote the eigenvalues associated with the same common eigenvector by $\gamma_l^{s_1}$ and $\gamma_l^{s_2}$, respectively, and set $\gamma_l^{s(t)}=\gamma_l^{s_1}$ during the first $\delta T$ units of each cycle and $\gamma_l^{s(t)}=\gamma_l^{s_2}$ during the remaining $(1-\delta)T$. The mode index therefore identifies the same common eigenvector in both phases; it need not order both spectra simultaneously. When the Laplacians do not share an eigenbasis, switching mixes the transverse modes and stability must instead be determined from the full $2(N-1)$-dimensional reduced monodromy matrix.
\par
\endgroup

\begingroup

With this modal decomposition, the full R\"ossler variational equations have the same form as \eqref{switching_full_variational_system}, with the coefficient $-2ds(t)$ replaced by $d\gamma_l^{s(t)}$. Explicitly, for $\mathbf{v}_l=(\xi_l,\eta_l,\zeta_l)^{\mathsf T}$,
\begin{equation}
    \dot{\mathbf{v}}_l=
    \begin{bmatrix}
        d\gamma_l^{s(t)}&-1&-1\\
        1&a&0\\
        z(t)&0&x(t)-c
    \end{bmatrix}\mathbf{v}_l,
    \quad l=2,\ldots,N,
    \label{variational system for complex network of switching rossler oscillators}
\end{equation}
where $(x(t),y(t),z(t))$ is the synchronous solution of \eqref{synchronous_solution_rossler}. Applying the same focal-plane approximation as in \eqref{reduced_s} gives
\begin{equation}
    \begin{aligned}
        \dot{\mathbf{u}}_l&=B_l^{s(t)}\mathbf{u}_l,
        \qquad \mathbf{u}_l=(\xi_l,\eta_l)^{\mathsf T},\\
        B_l^{s_i}&=\begin{bmatrix}d\gamma_l^{s_i}&-1\\1&a\end{bmatrix},
        \qquad i=1,2.
    \end{aligned}
    \label{variational system for complex network of switching reduced oscillators}
\end{equation}
Thus, network topology enters each reduced transverse system only through its two nonnegative effective modal couplings $-d\gamma_l^{s_1}$ and $-d\gamma_l^{s_2}$. In particular, on--off switching gives $\gamma_l^{s_1}=0$ and $\gamma_l^{s_2}=\gamma_l$, recovering \eqref{reduced_s} when $N=2$ and $\gamma_2=-2$.
\par
\endgroup

\subsection{Monodromy matrix and explicit stability conditions}
\begingroup

System~\eqref{variational system for complex network of switching reduced oscillators} is linear with constant coefficients within each switching phase. Its stability can therefore be examined using the same Floquet construction employed for the switched system in Ref.~\cite{porfiri2021antiresonance}. Set $\tau_1=\delta T$ and $\tau_2=(1-\delta)T$. Starting at the beginning of phase $s_1$, the evolution over one full period is
\begin{equation}
    \begin{aligned}
        \mathbf{u}_l(T)&=\Phi_l(T,0)\mathbf{u}_l(0),\\
        \Phi_l(T,0)&=e^{B_l^{s_2}\tau_2}e^{B_l^{s_1}\tau_1}.
    \end{aligned}
    \label{general period switching mapping on-off}
\end{equation}
The rightmost factor advances the perturbation through the first phase and the leftmost factor through the second. The product $\Phi_l(T,0)$ is the monodromy matrix. Repeating the cycle gives $\mathbf{u}_l(nT)=\Phi_l(T,0)^n\mathbf{u}_l(0)$, so its eigenvalues, the Floquet multipliers, determine whether perturbations decay from cycle to cycle. Equivalently, Floquet theory expresses the fundamental matrix as $\Phi_l(t,0)=\mathcal{P}_l(t)e^{R_lt}$, where $\mathcal{P}_l(t+T)=\mathcal{P}_l(t)$ and $R_l=T^{-1}\log\Phi_l(T,0)$ may be complex. Exponential stability is equivalent to negative real parts of both characteristic exponents, or to both Floquet multipliers lying strictly inside the unit circle.
\par
\endgroup

\begingroup
We need only the trace and determinant of this two-dimensional monodromy matrix to determine stability via the Jury criterion. Specifically, for the real matrix $\Phi_l=\Phi_l(T,0)$, both Floquet multipliers lie strictly inside the unit circle if and only if\cite{xiedeng2026dynamic}
\begin{equation}
    \det(\Phi_l)<1,
    \qquad
    \bigl|\operatorname{tr}(\Phi_l)\bigr|<1+\det(\Phi_l).
    \label{eq:jury_monodromy}
\end{equation}
We therefore compute these two scalar quantities directly, without introducing the eigenvectors of either frozen mode. For each $l$ and $i=1,2$, define
\begin{equation}
    \begin{aligned}
        h_{li}&=\frac{a+d\gamma_l^{s_i}}{2},\qquad
        q_{li}=\frac{a-d\gamma_l^{s_i}}{2},\\
        \kappa_{li}^2&=q_{li}^2-1,
    \end{aligned}
    \label{eq:phase_constants}
\end{equation}
and
\begin{equation}
    C_{li}=\cosh(\kappa_{li}\tau_i),\qquad
    S_{li}=\begin{cases}
        \sinh(\kappa_{li}\tau_i)/\kappa_{li},&\kappa_{li}\ne0,\\
        \tau_i,&\kappa_{li}=0.
    \end{cases}
    \label{eq:phase_functions}
\end{equation}
These quantities are real even when $\kappa_{li}$ is imaginary: if $\kappa_{li}=\mathrm{i}\omega_{li}$, they become $C_{li}=\cos(\omega_{li}\tau_i)$ and $S_{li}=\sin(\omega_{li}\tau_i)/\omega_{li}$. Since $(B_l^{s_i}-h_{li}I_2)^2=\kappa_{li}^2 I_2$, the phase transition matrix is
\begin{equation}
    e^{B_l^{s_i}\tau_i}=e^{h_{li}\tau_i}
    \begin{bmatrix}
        C_{li}-q_{li}S_{li}&-S_{li}\\
        S_{li}&C_{li}+q_{li}S_{li}
    \end{bmatrix}.
    \label{eq:phase_exponential}
\end{equation}
The limiting definition in \eqref{eq:phase_functions} also covers a repeated eigenvalue of a frozen mode.
\par
\endgroup

\begingroup
Let
\begin{equation}
    \begin{aligned}
        \bar\gamma_l&=\delta\gamma_l^{s_1}+(1-\delta)\gamma_l^{s_2},\\
        E_l&=\exp\!\left[\frac{1}{2}(a+d\bar\gamma_l)T\right].
    \end{aligned}
    \label{eq:cycle_constants}
\end{equation}
Multiplication of the two matrices in \eqref{eq:phase_exponential} yields
\begin{equation}
    \begin{aligned}
        \chi_l:=\operatorname{tr}\Phi_l
        &=2E_l\big[C_{l1}C_{l2}\\
        &\qquad +(q_{l1}q_{l2}-1)S_{l1}S_{l2}\big],\\
        D_l:=\det\Phi_l&=E_l^2=e^{(a+d\bar\gamma_l)T}.
    \end{aligned}
    \label{eq:monodromy_invariants}
\end{equation}
The determinant also follows immediately from $\det(e^B)=e^{\operatorname{tr}B}$. The Floquet multipliers are consequently
\begin{equation}
    \varphi_{l,\pm}=\frac{\chi_l\pm\sqrt{\chi_l^2-4D_l}}{2}.
    \label{eq:floquet_multipliers}
\end{equation}
Substituting the trace and determinant from \eqref{eq:monodromy_invariants} into the Jury conditions \eqref{eq:jury_monodromy}, we arrive at the following assertion.
\begin{Assertion}[Stability criterion for the focal-plane approximation]
\itshape
\label{statement:reduced_synchronization}
Consider the periodically switching network \eqref{model} of identical
$x$-coupled R\"ossler oscillators, with $T>0$, $0<\delta<1$, $d>0$,
and symmetric negative-semidefinite Laplacians $L^{s_1}$ and $L^{s_2}$
that share a common orthonormal eigenbasis, equivalently, that commute:
$L^{s_1}L^{s_2}=L^{s_2}L^{s_1}$. In the focal-plane approximation
\eqref{variational system for complex network of switching reduced oscillators},
complete synchronization is linearly exponentially stable if and only if,
for every transverse mode $l=2,\ldots,N$,
\begin{equation}
    \begin{aligned}
        1-\chi_l+D_l&>0,\;
        1+\chi_l+D_l>0,\\
        D_l&<1
        \quad\Longleftrightarrow\quad -d\bar\gamma_l>a,
    \end{aligned}
    \label{eq:reduced_sync_criterion}
\end{equation}
where $\chi_l$ and $D_l$ are given by
\eqref{eq:monodromy_invariants}, with the constants defined in
\eqref{eq:phase_constants}--\eqref{eq:cycle_constants}.
Equivalently, $-d\bar\gamma_l>a$ and $|\chi_l|<1+D_l$ must hold
for every transverse mode. These conditions apply to on-off switching
for any fixed undirected topology by setting $\gamma_l^{s_1}=0$
and $\gamma_l^{s_2}=\gamma_l$.
\end{Assertion}
\par
\endgroup

\begingroup
\begin{proof}
The common eigenbasis splits the reduced transverse dynamics into
the $N-1$ independent systems
\eqref{variational system for complex network of switching reduced oscillators}.
For each system, Floquet theory gives uniform exponential decay
if and only if both eigenvalues of
\eqref{general period switching mapping on-off} have modulus less
than one. Applying the quadratic Jury criterion gives the first
two inequalities in \eqref{eq:reduced_sync_criterion}.
Since $\operatorname{tr}(B_l^{s_i})=a+d\gamma_l^{s_i}$,
the determinant is $D_l=\exp[(a+d\bar\gamma_l)T]$.
For $T>0$, the remaining inequality $D_l<1$ is therefore
equivalent to $a+d\bar\gamma_l<0$, or $-d\bar\gamma_l>a$.
Because there are finitely many transverse modes, all transverse
perturbations decay exponentially exactly when the criterion
holds for every mode.
\end{proof}
\par
\endgroup

The inequalities are strict: equality identifies a possible stability boundary, rather than exponential decay. At $\chi_l=1+D_l$ or $\chi_l=-(1+D_l)$, a multiplier equals $+1$ or $-1$, respectively. The remaining boundary is $D_l=1$ with $|\chi_l|\leq2$. The determinant condition requires net area contraction over a cycle, but is not sufficient by itself; the trace condition also accounts for the relative orientation of expansion and contraction during the two phases.

Assertion~\ref{statement:reduced_synchronization} concerns the reduced transverse equations. For the original nonlinear network, local exponential attraction to the synchronization manifold requires uniform exponential stability of the full transverse variational dynamics along that manifold. Since the focal-plane approximation omits the $\zeta_l$ dynamics and its feedback, the reduced criterion alone does not establish this property for the full network. It instead provides explicit analytical predictions that we compare below with the largest transverse Lyapunov exponent of \eqref{variational system for complex network of switching rossler oscillators}.

\subsection{Master stability function and windows of opportunity}
For on--off switching, all dependence on the active topology enters through the nonnegative coupling magnitude $-d\gamma_l$. With $a$ and $\delta$ fixed, we define the reduced master stability function by
\begin{equation}
    \mathcal{L}_{\rm red}(-d\gamma_l,T)
    =\frac{1}{T}\ln\rho\!\left(\Phi(T,0;0,-d\gamma_l)\right),
    \label{eq:reduced_msf}
\end{equation}
where $\rho$ is the spectral radius and the last two arguments specify the effective couplings in the off and on phases. This is the largest real part of the Floquet exponents. In analogy with the static master stability approach, \cite{pecora1998master} synchronization is linearly stable in the reduced description precisely when
\begin{equation}
    \max_{l=2,\ldots,N}\mathcal{L}_{\rm red}(-d\gamma_l,T)<0.
    \label{eq:network_msf_criterion}
\end{equation}
The switching period is thus an additional control parameter of the master stability function.

\begin{figure}[!tbp]
    \centering
    \includegraphics[width=\linewidth]{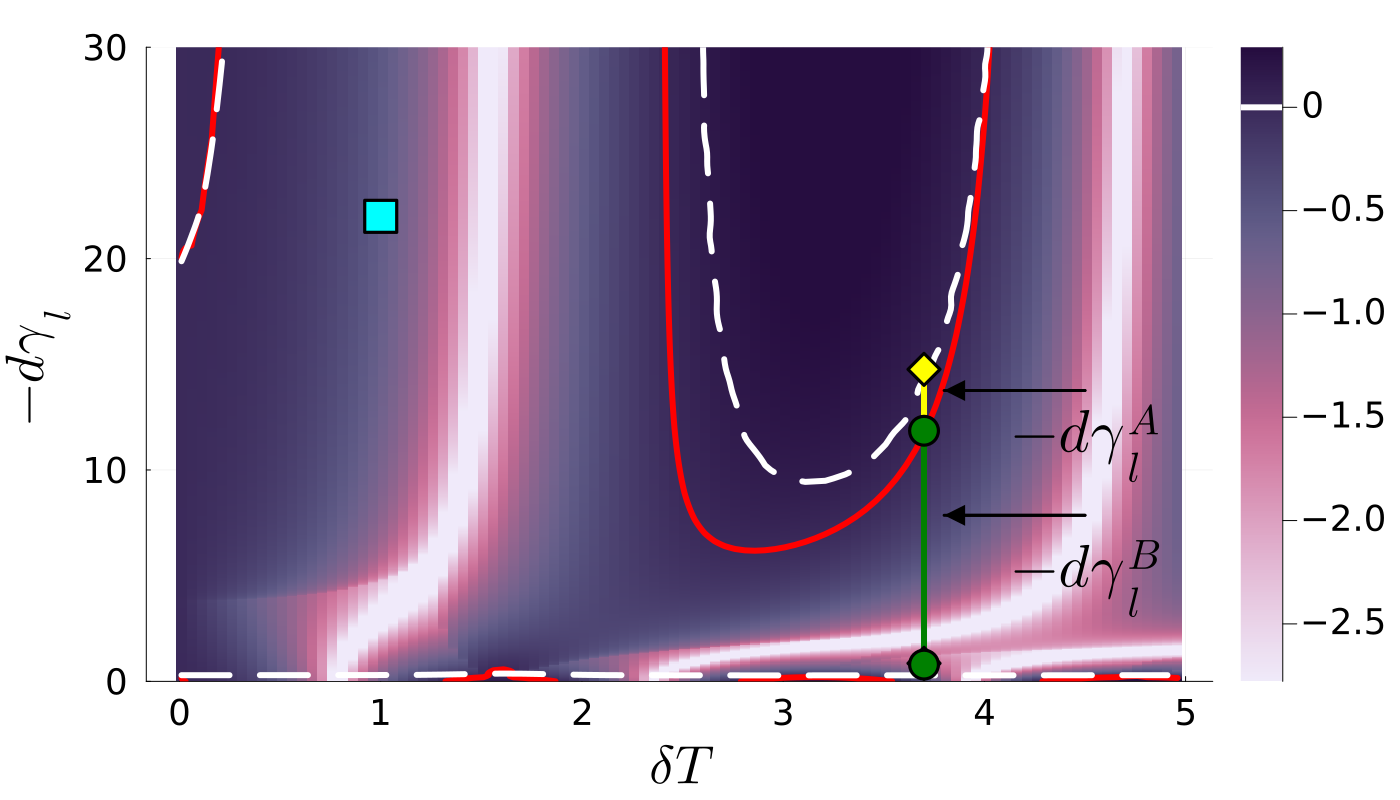}
    \caption{Master stability diagram for periodic on--off switching, with $a=0.1$ and $\delta=1/2$. The horizontal coordinate is the dwell time $\delta T=T/2$; {the vertical coordinate is the effective modal coupling $-d\gamma_l$.} Color shows the logarithmic growth per cycle, $\ln\rho(\Phi_l)=T\mathcal{L}_{\rm red}(-d\gamma_l,T)$; negative values indicate transverse decay. The red curve is the reduced analytical zero contour, and the white dashed curve is the numerically computed zero contour of the largest transverse Lyapunov exponent of \eqref{variational system for complex network of switching rossler oscillators}. {The cyan square at $(\delta T,-d\gamma_l)=(1.0,22)$ locates the two-node example of Fig.~\ref{Switching Period Comparison}.} At $\delta T=3.7$ and $d=1.2$, the yellow and green segments span the effective couplings of networks A and B introduced in Fig.~\ref{Pruning algorithm stability}; endpoint markers indicate $-d\gamma_2$ and $-d\gamma_{50}$.}
    \label{Analytical stability region for on-off switching for general network.}
\end{figure}

Figure~\ref{Analytical stability region for on-off switching for general network.} can be read in two complementary ways. A horizontal slice fixes $-d\gamma_l$ and identifies the switching intervals for which that transverse mode contracts. For $-d\gamma_l=22$, the reduced model predicts the first two stable intervals as approximately $\delta T\in(0.104,2.421)$ and $(3.955,5.290)$. The cyan square lies in the first interval and recovers the two-node window of opportunity of Sec.~\ref{sec:two_node}. In contrast, the fast-switching limit is governed by the average reduced matrix, which is stable only when $a<-(1-\delta)d\gamma_l<1/a$ for $a>0$. At $a=0.1$, $\delta=1/2$, and $-d\gamma_l=22$, the average effective coupling is $11>1/a=10$, so the fast-switching limit is unstable. The bounded intervals of negative master stability therefore quantify the stabilization from favorable switching times, even though the average system is unstable.

\begingroup

A vertical slice fixes the switching schedule and identifies the effective modal couplings $-d\gamma_l$ for which transverse perturbations decay. Every transverse mode must lie in this stable set. When the set is a single interval, this is equivalent to requiring that the spectral span $[-d\gamma_2,-d\gamma_N]$ be contained in it; when the set has several components, each eigenvalue must be checked individually. This interpretation turns the master stability diagram into a guide for changing network topology while keeping the coupling strength and switching schedule fixed.

\subsection{Interplay between network topology and switching rate}
\label{sec:topology_switching}
The yellow spectral span in Fig.~\ref{Analytical stability region for on-off switching for general network.} illustrates a network whose strongest effective coupling exceeds the upper stability boundary at $\delta T=3.7$. To induce synchronization at this fixed switching interval, its spectrum must move downward into the stable band. Among edge addition and edge removal, only removal moves the spectrum in the required direction for this overcoupled network. This conclusion follows from the location of its spectrum relative to the stability region; it is not a general prohibition on synchronization by edge addition.

With our negative-semidefinite Laplacian convention, deleting an unweighted edge $e=(i,j)$ gives $L_{\rm new}=L_{\rm old}+\mathbf{b}_e\mathbf{b}_e^{\mathsf T}$, where $\mathbf{b}_e=\mathbf{e}_i-\mathbf{e}_j$ and $\mathbf{e}_i$ is the $i$th coordinate vector. Hence every ordered signed eigenvalue $\gamma_l$ is nondecreasing, while its magnitude $-\gamma_l$ and the effective modal coupling $-d\gamma_l$ are nonincreasing. Edge pruning can therefore bring the upper endpoint into the stable band. At the same time, it lowers the smallest effective coupling, so excessive pruning can push that endpoint below the lower boundary and destroy synchronization again.

We demonstrate this effect using an Erd\H{o}s--R\'{e}nyi network of $N=50$ R\"ossler oscillators with $120$ edges, denoted by A in Fig.~\ref{Pruning algorithm stability}. Throughout the experiment, $d=1.2$, $\delta=1/2$, and $\delta T=3.7$ are fixed. {We use periodic on--off switching, with $L^{s_1}=\mathbf{0}_{N\times N}$ and $L^{s_2}=L.$ The network is uncoupled for $\delta T$ units of each cycle and coupled through that topology for the remaining $(1-\delta)T$.} The original network has $-\gamma_2^{A}\simeq0.71$ and $-\gamma_{50}^{A}\simeq12.30$, giving the yellow spectral span in Fig.~\ref{Analytical stability region for on-off switching for general network.}. Its upper endpoint, $-d\gamma_{50}^{A}\simeq14.76$, lies above both the reduced analytical boundary and the full-system numerical boundary.

Algorithm~\ref{Minimal spectral radius edge pruning algorithm}, given in Appendix~\ref{Topology adaptation algorithms}, constructs a sequence of connected graphs by removing one edge at a time. Among all candidate deletions that preserve connectivity, it selects the one that minimizes the spectral radius of the resulting Laplacian, $\rho(L)=-\gamma_N$. This greedy rule targets the excessive upper endpoint. It also records the lower endpoint, $-\gamma_2$, because connectivity alone does not ensure it remains within the synchronization interval.

After nine removals, network B has $111$ edges, $-\gamma_2^{B}\simeq0.66$, and $-\gamma_{50}^{B}\simeq9.88$. Its spectral span is shown in green in Fig.~\ref{Analytical stability region for on-off switching for general network.}. The full network now synchronizes, as demonstrated in Fig.~\ref{Pruning algorithm stability}(c). The reduction predicts the direction of this change, although its upper stability threshold, approximately $11.72$ at $\delta T=3.7$, is slightly below the endpoint $-d\gamma_{50}^{B}\simeq11.86$. Thus, the numerical onset of synchronization precedes the reduced prediction in this example. We assess synchronization directly using
the synchronization error $\Delta_{\rm sync}$ defined in \eqref{eq:network_sync_error}.
The continued pruning sweep in Fig.~\ref{Pruning algorithm stability}(a) reveals a bounded range of successful modifications: synchronization is observed after $2$ through $43$ removals and is lost at the $44$th removal. Network C, obtained after $50$ removals, has $70$ edges and $-d\gamma_2^{C}\simeq0.17$, below the reduced lower boundary $0.2$. Its failure to synchronize illustrates the obstruction at the opposite end of the spectrum.

\begin{figure}[!tbp]
    \centering
    {(a)}\\
    \includegraphics[width=\linewidth]{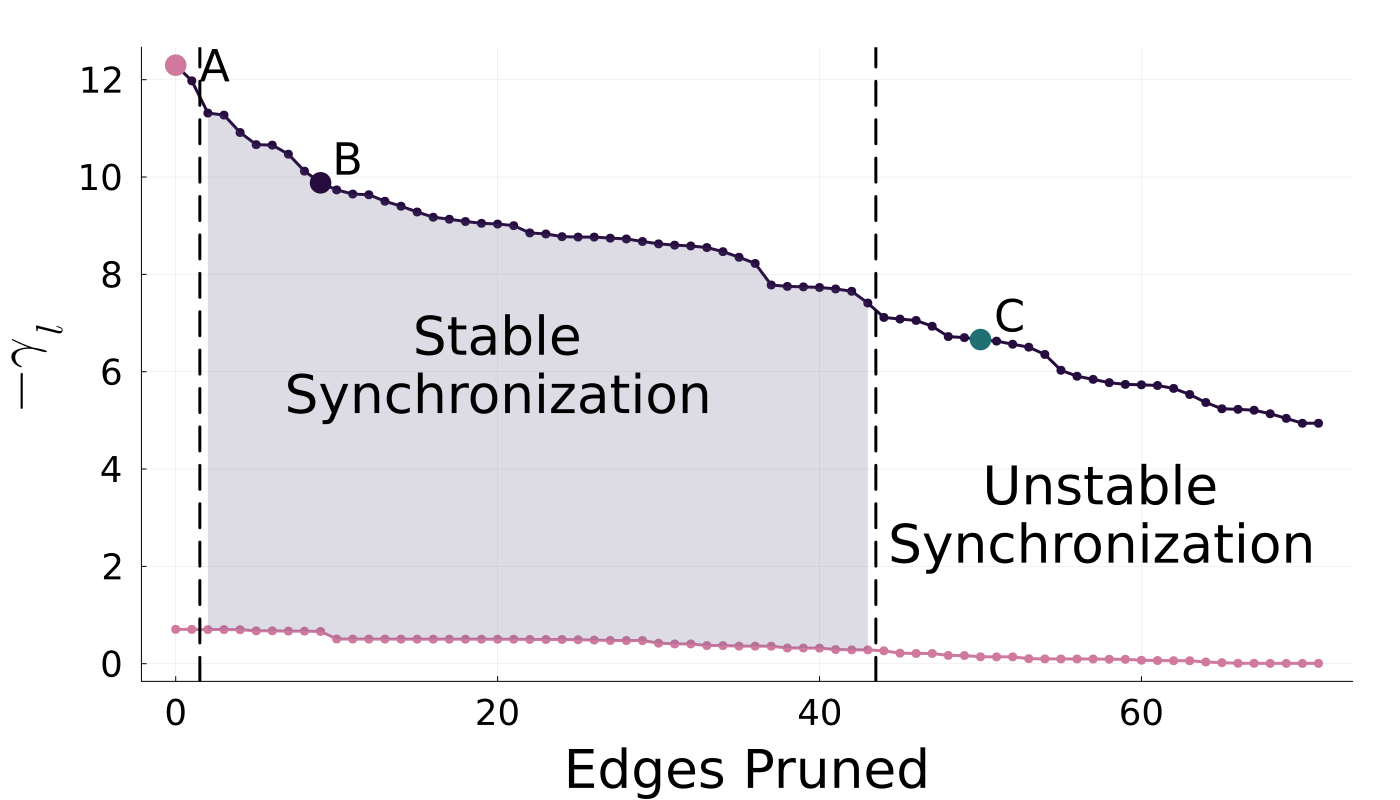}
    {(b)}\\
    \includegraphics[width=\linewidth]{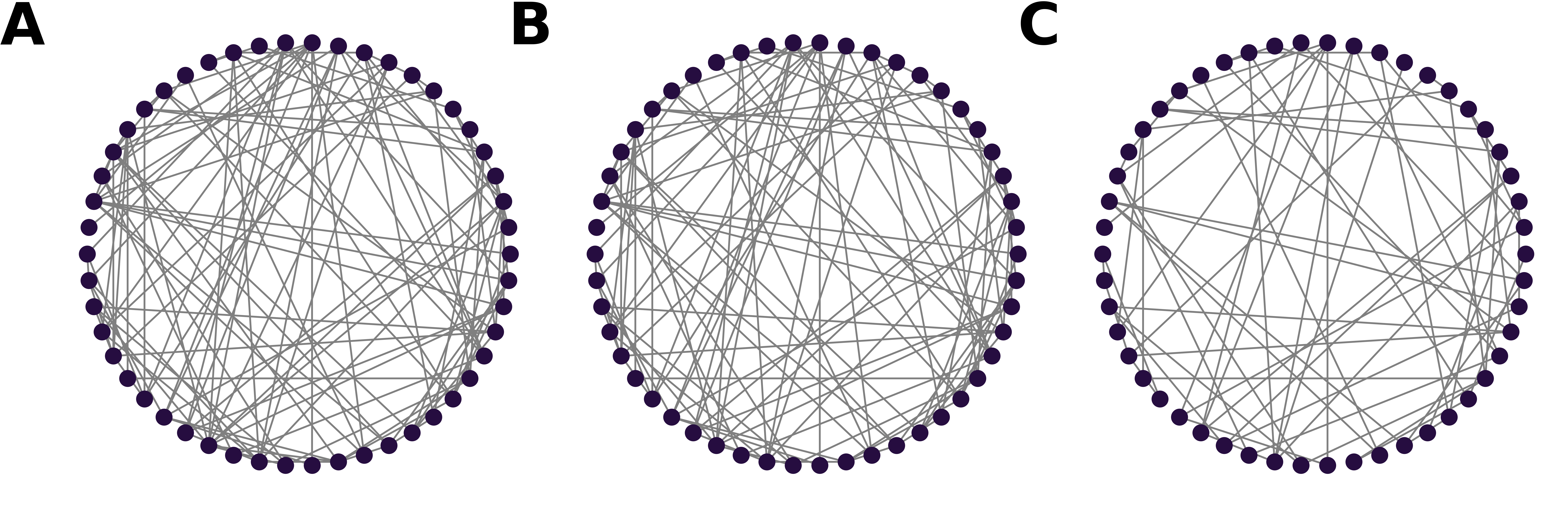}
    {(c)}\\
    \includegraphics[width=\linewidth]{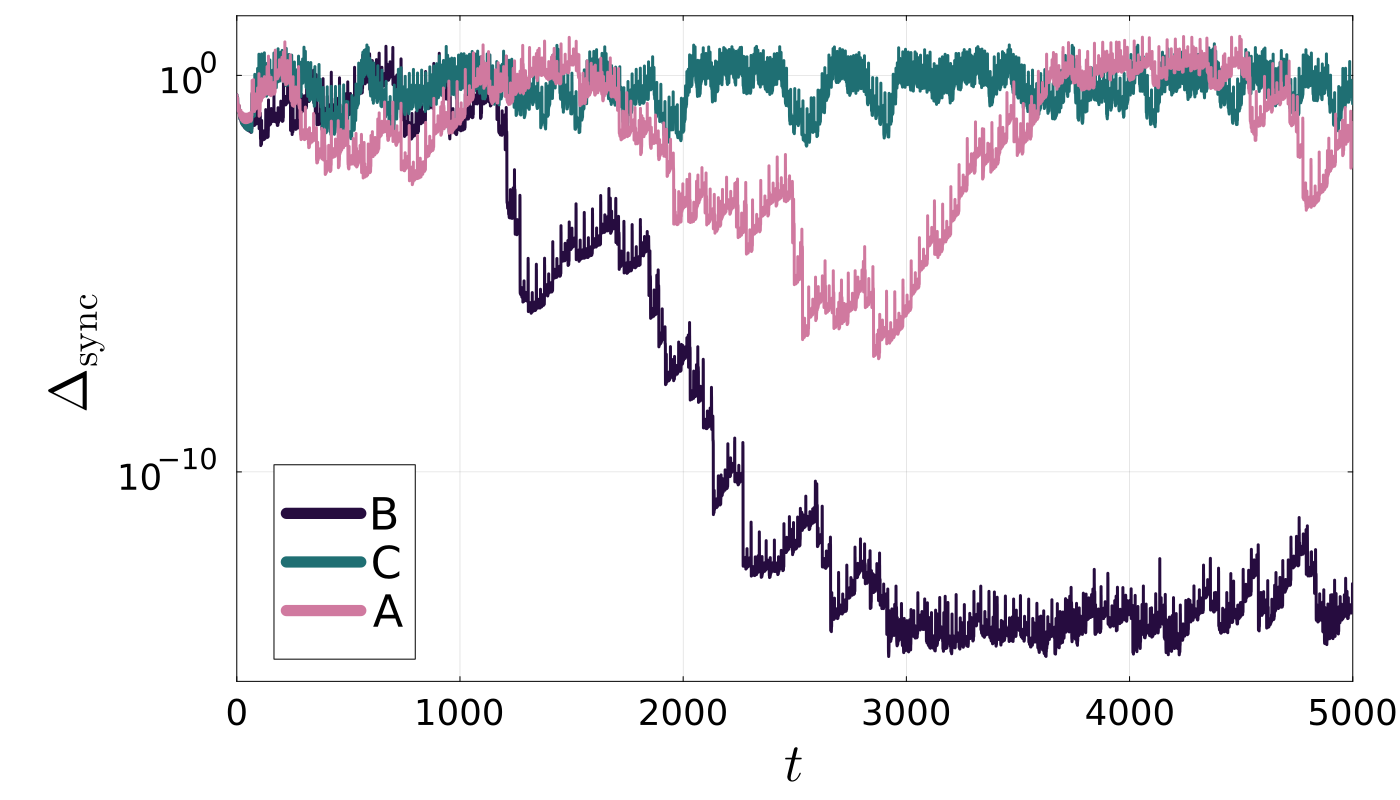}
    \caption{{Synchronization induced and then lost through edge pruning in a $50$-node R\"ossler network under periodic on--off switching, at fixed $d=1.2$, $\delta=1/2$, and dwell time $\delta T=3.7$.} (a) Extreme Laplacian magnitudes $-\gamma_{50}$ (dark purple) and $-\gamma_2$ (pink) along Algorithm~\ref{Minimal spectral radius edge pruning algorithm}. The shaded range, $2$--$43$ removals, is identified by the full-system numerical results; it is not an exact stability guarantee from the reduced model. (b) Representative graphs: A, the original network ($120$ edges); B, after $9$ removals ($111$ edges); and C, after $50$ removals ($70$ edges). (c) Synchronization error \eqref{eq:network_sync_error} for these graphs. Only B synchronizes. Network A violates the upper modal-coupling bound, whereas excessive pruning leaves network C below the lower bound.}
    \label{Pruning algorithm stability}
\end{figure}

This experiment makes the interplay between timing and topology explicit. The switching interval selects the stable set of effective couplings, and the graph spectrum determines whether every transverse mode belongs to that set. Removing edges can consequently restore synchronization without changing either the switching schedule or the strength of any surviving edge. The useful direction of topology modification depends on the shape of the selected stable set. We next show that the same focal-plane analysis can instead identify a regime in which adding edges induces synchronization.

\section{Beyond R\"ossler oscillators: Windows of Opportunity and Edge Augmentation}
\label{sec:piecewise_main}
The focal-plane reduction and its prediction of windows of opportunity
extend beyond R\"ossler dynamics to networks of other saddle-focus
oscillators. We illustrate this broader applicability using the
piecewise-linear oscillator studied in
Refs.~\cite{barajas2021generating,tang2022multilayer}, whose network
master stability diagram also reveals a complementary route to
synchronization through edge augmentation. Its on-off coupled network is
\begin{equation}
    \dot{\mathbf{x}}_i=\mathbf{F}_{\rm pw}(\mathbf{x}_i)
       +d s(t)\sum_{j=1}^{N}L_{ij}P\mathbf{x}_j,
    \label{piecewise system}
\end{equation}
where $\mathbf{x}_i=(x_i,y_i,z_i)^{\mathsf T}$ and $P=\operatorname{diag}(1,0,0)$. {The coupling follows a periodic on--off protocol: $s(t)=s_1=0$ during the first $\delta T$ units of each cycle and $s(t)=s_2=1$ during the remaining $(1-\delta)T$. Thus, $L^{s_1}=\mathbf{0}_{N\times N}$ and $L^{s_2}=L$.} The node dynamics is
\begin{equation}
    \mathbf{F}_{\rm pw}(\mathbf{x})=
    \begin{cases}
        M_{\rm f}\mathbf{x},&y+z<1,\\
        M_{\rm r}(\mathbf{x}-\mathbf{b}),&y+z\geq1,
    \end{cases}
    \label{piecewise subsystems}
\end{equation}
with
\begin{equation}
    \begin{aligned}
        M_{\rm f}&=\begin{bmatrix}q&-1&0\\1&q&0\\0&0&-2\end{bmatrix},
        &M_{\rm r}&=\begin{bmatrix}-1.5&0&0\\0&0&-1\\0&1&0\end{bmatrix},\\
        \mathbf{b}&=(0,1/2,1/2)^{\mathsf T},&q&=0.2.
    \end{aligned}
    \label{eq:piecewise_node_matrices}
\end{equation}
The focal part has eigenvalues $q\pm\mathrm{i}$ and $-2$, combining planar expansion and rotation with contraction in the third direction. The other branch returns the trajectory toward the focal region. Here, switching between the two branches of the individual oscillator is state dependent and is distinct from the externally prescribed on--off coupling.

\subsection{Reduced stability diagram and windows of opportunity}
Retaining the focal dynamics gives the reduced transverse system
\begin{equation}
    \begin{aligned}
        \dot{\mathbf{u}}_l&=B_{{\rm pw},l}^{s(t)}\mathbf{u}_l,\\
        B_{{\rm pw},l}^{s(t)}&=
        \begin{bmatrix}q+d s(t)\gamma_l&-1\\1&q\end{bmatrix},
        \quad l=2,\ldots,N.
    \end{aligned}
    \label{two node piecewise reduced variational system}
\end{equation}
As in the R\"ossler example, the coefficients are constant within each coupling phase. The monodromy construction and the Jury criterion therefore apply directly, with the trace and determinant given in Appendix~\ref{app:piecewise}. The reduced master stability function is denoted by $\mathcal{L}_{{\rm pw},{\rm red}}(-d\gamma_l,T)$. We also retain the comparison with static coupling and the focal-region diagnostics in that appendix, allowing us to focus here on the switching windows and their implications for topology adaptation.

Figure~\ref{Analytical stability region for on-off switching in two node system.}(a) shows the resulting master stability diagram. Horizontal slices again contain alternating stable and unstable intervals. At $-d\gamma_l=4$, Figs.~\ref{Analytical stability region for on-off switching in two node system.}(a)-(b) compare the reduced prediction with the full-system numerical results: synchronization is stable for the selected short dwell time, is lost in an intermediate band, and is restored in a second window near $\delta T=4.2$. The recurrence of such windows in two different oscillator models supports the role of the focal geometry in the reduction. Their quantitative boundaries differ, so the reduced conditions remain predictions for the original networks rather than exact nonlinear synchronization criteria.

\begin{figure}[!tbp]
    \centering
    \includegraphics[width=\linewidth]{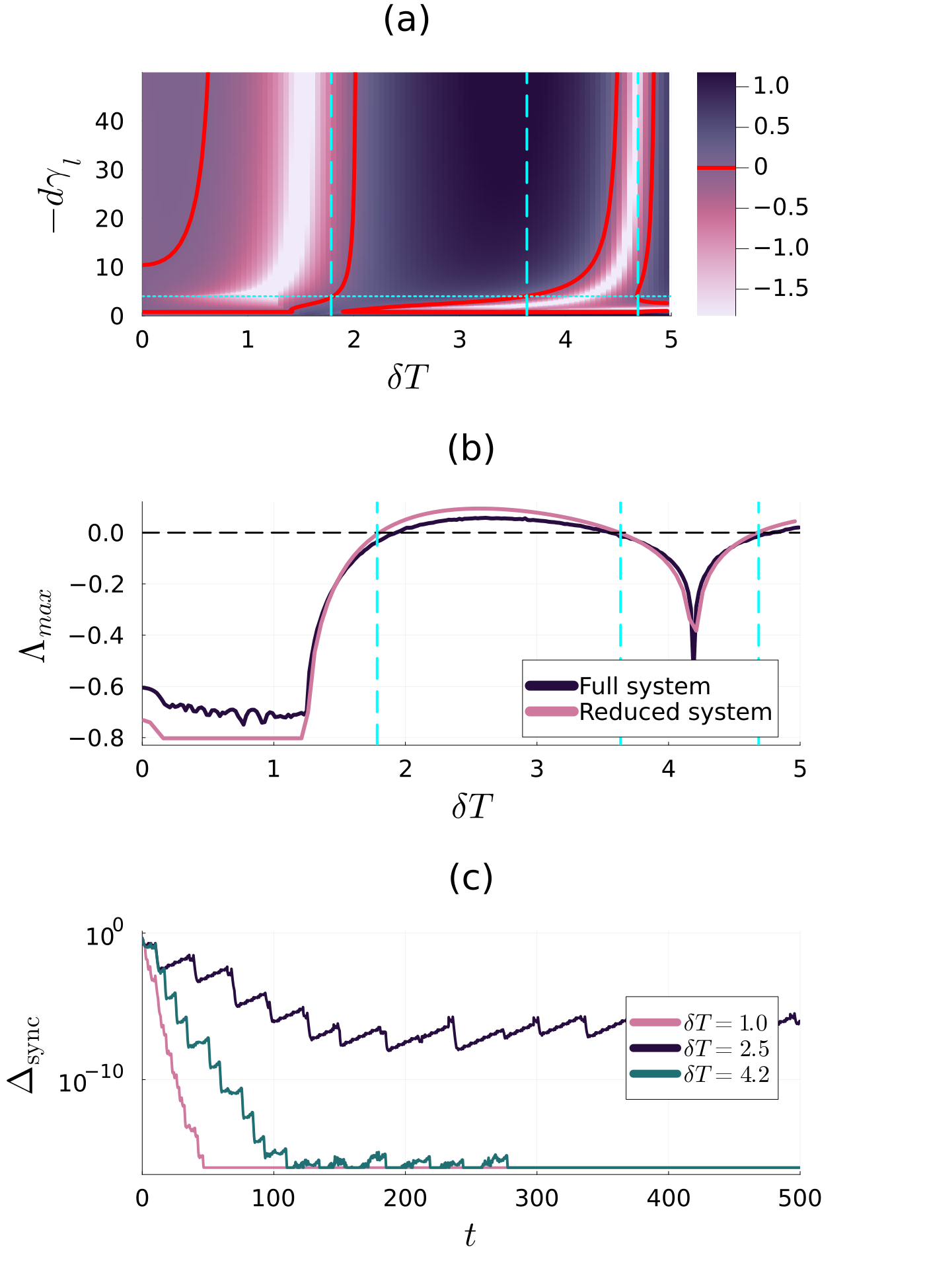}
    \caption{{Windows of opportunity in the piecewise-linear network \eqref{piecewise system} under periodic on--off switching, with $q=0.2$ and $\delta=1/2$.
(a) Reduced master stability diagram: color shows $\ln\rho(\Phi_{{\rm pw},l})$, and red contours mark zero growth. The cyan horizontal line is $-d\gamma_l=4$, corresponding to the two-node network with $d=2$ considered in panels (b) and (c).
(b) Rate-selective transverse stability of this two-node network: the exponent $T^{-1}\ln\rho(\Phi_{{\rm pw},l})$ of the reduced system (pink) is compared with the full-system numerical exponent (dark purple). The analytical crossings occur near $\delta T=1.8$, $3.6$, and $4.7$.
(c) Corresponding synchronization error $\Delta_{\rm sync}$, defined in \eqref{eq:network_sync_error}, for representative dwell times in the successive stable and unstable bands. The errors at $\delta T=1$ and $4.2$ decay to numerical precision; at $\delta T=2.5$, the small residual error persists and grows after a transient. The selected trajectories illustrate the stability changes predicted by the diagram. }}
    \label{Analytical stability region for on-off switching in two node system.}
\end{figure}

\subsection{Inducing synchronization by adding edges}
The vertical-slice interpretation now suggests a topology modification complementary to pruning. At $\delta T=1.75$, the reduced diagram in Fig.~\ref{Analytical stability region for on-off switching in two node system.}(a) has a lower stability threshold $-d\gamma_l\simeq3.4275$ and no upper boundary in the coupling range relevant to the following experiment. For $N=20$, $d=1.2$, and unweighted graphs, all effective couplings satisfy $-d\gamma_l\leq dN=24$; throughout this range, the numerically evaluated reduced synchronization criterion is
\begin{equation}
    -d\gamma_2>3.4275.
    \label{augmentation criterion}
\end{equation}
All other transverse eigenvalues then lie in the same stable interval. Thus, the obstruction is insufficient effective coupling at the lower end of the spectrum, rather than excessive coupling at the upper end as in the R\"ossler pruning example.

Adding an edge changes the negative Laplacian according to $L_{\rm new}=L_{\rm old}-\mathbf{b}_e\mathbf{b}_e^{\mathsf T}$, so the magnitudes $-\gamma_l$ are nondecreasing. Edge augmentation can therefore move an initially undercoupled network across the threshold in \eqref{augmentation criterion}.  Algorithm~\ref{Minimal spectral radius edge augmenting algorithm} in Appendix~\ref{Topology adaptation algorithms} starts from a path on $20$ nodes and adds one missing edge at each step, choosing the candidate that minimizes the resulting spectral radius. This choice provides a reproducible sequence for tracking the motion of both spectral endpoints; it does not optimize the number of edges needed to synchronize.

The on--off switching schedule remains fixed throughout the augmentation experiment; only the topology active during the on phase is modified between successive simulations.

Figure~\ref{Edge augmenting stability} follows this sequence from the path graph ($19$ edges) to the complete graph ($190$ edges), requiring $171$ additions. The lower spectral endpoint crosses the reduced threshold $-\gamma_2>3.4275/1.2\simeq2.8563$ after $69$ additions. In the full-network simulations, synchronization appears earlier, after $53$ additions, and persists along the rest of this sequence. The reduction is therefore conservative at the onset in this numerical experiment. Panel (c) contrasts the initial graph A with two augmented graphs B and C: the initial graph fails to synchronize, whereas both augmented graphs reduce the error by many orders of magnitude.

\begin{figure}[!tbp]
    \centering
    {(a)}\\
    \includegraphics[width=\linewidth]{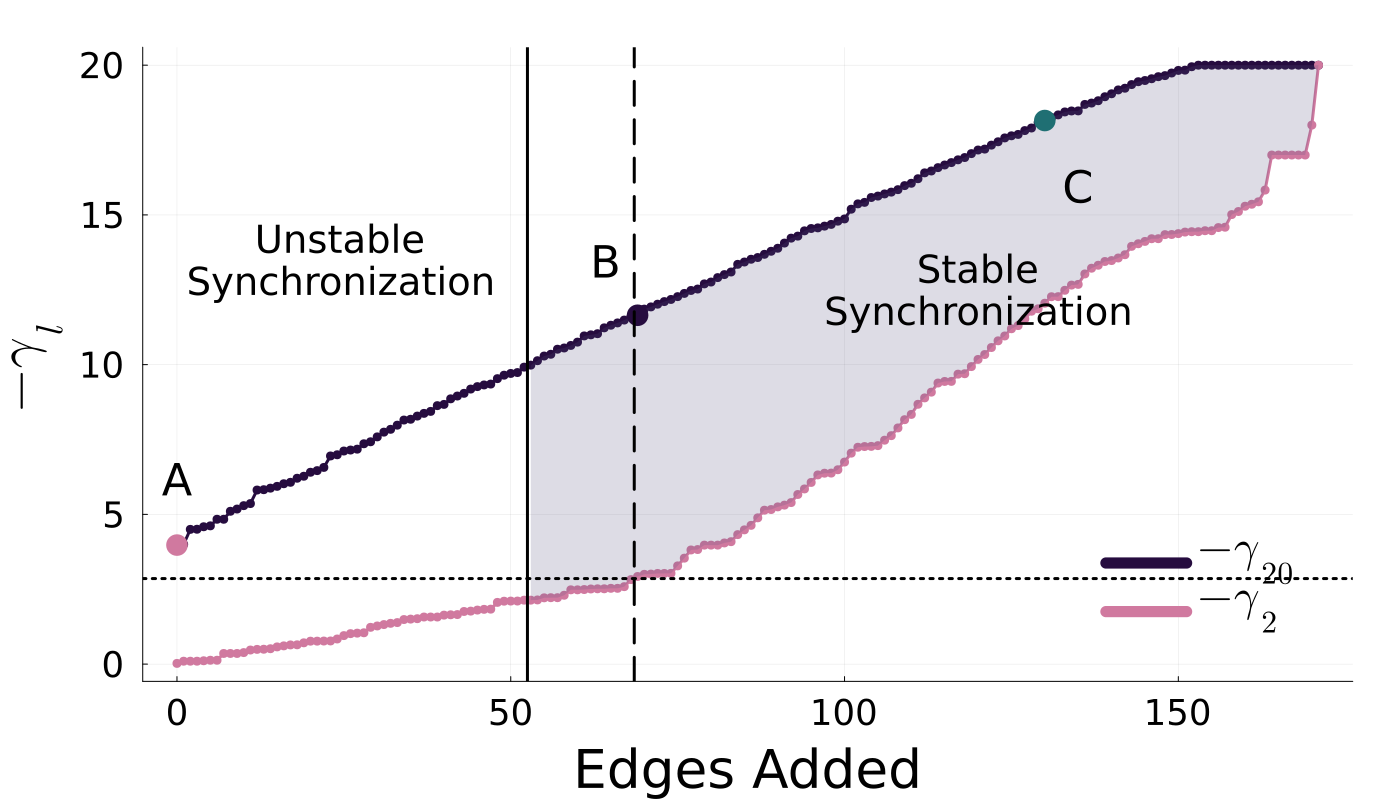}
    {(b)}\\
    \includegraphics[width=\linewidth]{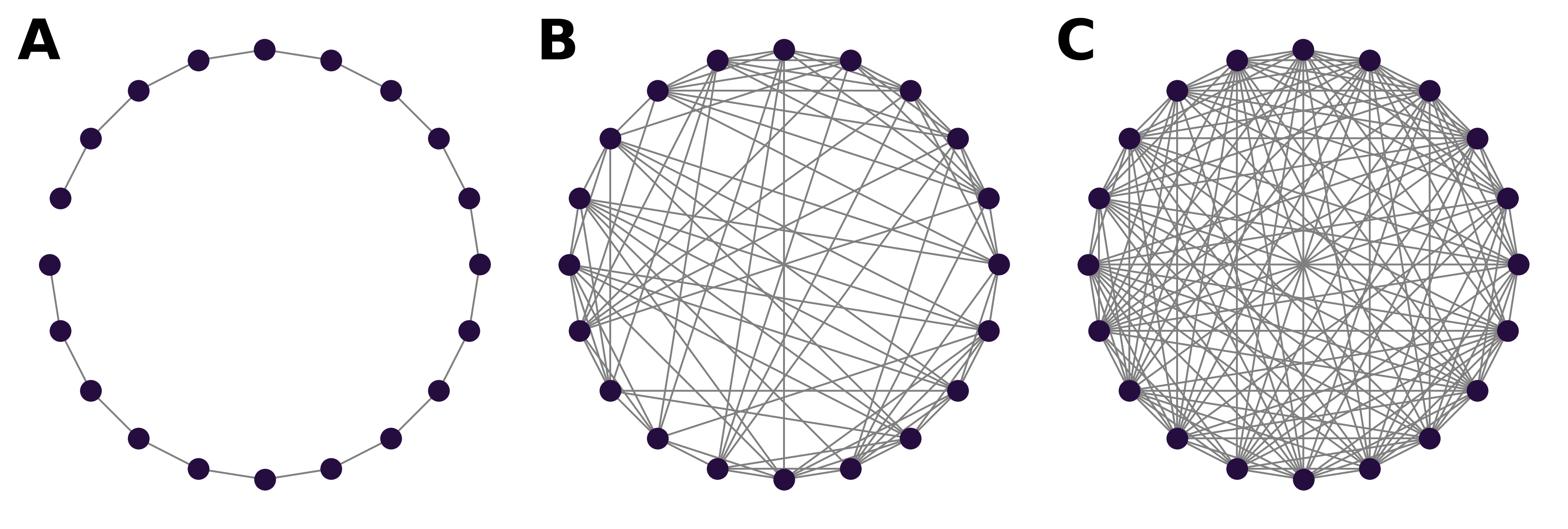}
    {(c)}\\
    \includegraphics[width=\linewidth]{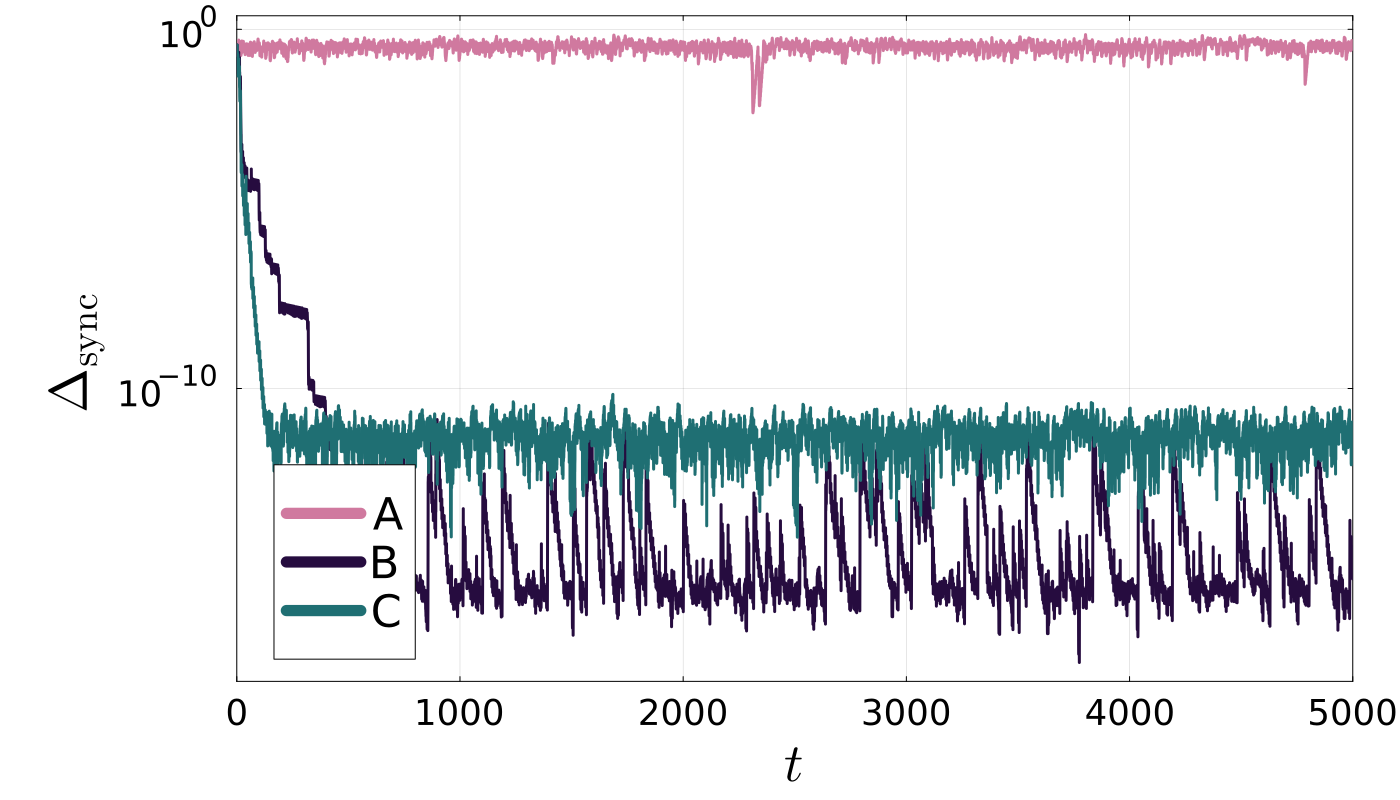}
    \caption{{Synchronization induced by edge augmentation in a $20$-node piecewise-linear network under periodic on--off switching, at fixed $d=1.2$, $\delta=1/2$, and dwell time $\delta T=1.75$.} (a) Extreme Laplacian magnitudes $-\gamma_{20}$ (dark purple) and $-\gamma_2$ (pink) along Algorithm~\ref{Minimal spectral radius edge augmenting algorithm}. The dotted horizontal line is the reduced threshold $-\gamma_2\simeq2.8563$. The dashed vertical line marks its crossing after $69$ additions, whereas the solid vertical line and shading indicate synchronization in the full-network simulations from $53$ additions onward. (b) Graphs A, B, and C have $19$, $88$, and $149$ edges, respectively. (c) Their synchronization errors \eqref{eq:network_sync_error}. Each of networks B and C exhibits stable synchronization, whereas the network with the initial path topology A
does not. This example complements Fig.~\ref{Pruning algorithm stability}: at the selected switching schedule, adding edges moves the spectrum into the stable set.}
    \label{Edge augmenting stability}
\end{figure}

Together, the two examples show why the effect of network connectivity cannot be separated from the switching schedule. A bounded stable interval can make edge removal beneficial for an overcoupled graph, whereas a stable interval extending to stronger coupling can make edge addition beneficial for an undercoupled graph. The same reduced Floquet calculation identifies both possibilities. These conclusions concern periodic switching; we next examine which features persist when the coupling schedule is randomized.

\section{Periodic Versus Stochastic Switching}
\label{sec:stochastic_switching}
Periodicity lets us represent each switching cycle with the same monodromy matrix and derive exact stability conditions for the reduced system. Under stochastic switching, perturbations instead evolve through products of transition matrices whose order or dwell times depend on the realization. Consequently, a negative periodic master stability function does not by itself establish stability for a random schedule, even when the two schedules have the same mean coupling. The numerical comparison below examines this distinction for both oscillator models.

We compare periodic synchronization with the fraction of successful runs among $1,000$ independent stochastic switching realizations at each plotted interval. The numerical classification uses a synchronization-error tolerance of $10^{-4}$ and a finite observation interval.
{Both comparisons use the same on-phase topology, a connected $50$-node Erd\H{o}s--R\'{e}nyi graph with $109$ edges, $-\gamma_2\simeq0.52$, and $-\gamma_{50}\simeq11.22$, which is distinct from network A of Fig.~\ref{Pruning algorithm stability}. All coupling is absent in the off phase. Under the periodic schedule, the network alternates between the two phases every $\delta T$, beginning in the off phase. Under the stochastic schedule, time is divided into consecutive intervals of length $\delta T$, and the coupling is switched on during each interval independently with probability $1/2$. The dwell time is fixed; only the mode sequence is random, and both schedules have a mean coupling of $d/2$. Each of the $3N$ initial state variables is drawn independently and uniformly from $[0,1]$. The equations are integrated with the Runge--Kutta--Fehlberg 7(8) scheme at the fixed step $\min(\delta T,0.1)$.
A run is classified as unsynchronized if $\sum_{i}(|x_i|+|y_i|+|z_i|)$ exceeds $500N$ at the end of any interval. {Otherwise, synchronization is assessed using the error $\Delta_{\rm sync}$ defined in \eqref{eq:network_sync_error}, evaluated at the final integration time. A run is counted as synchronized when this terminal error is below $10^{-4}$.} For the R\"ossler network, $d=3$ and the final time is $10^4$. The periodic schedule is simulated once at each {$\delta T\in\{0.01,0.11,0.21,\ldots,9.91\}$}, and the stochastic success fraction is estimated from $1,000$ realizations on a grid in $\delta T$ with spacing $0.025$. For the piecewise-linear network, $d=2$ and the final time is $2,000$. Both schedules use the spacing $0.025$, and the initial states and switching sequences are generated from fixed random seeds.
}

For the $50$-node R\"ossler network at $d=3$, Fig.~\ref{Stochastic vs periodic rossler} shows a broad principal window, approximately $\delta T\in(0.31,2.51)$, where periodic switching succeeds, and the stochastic success probability is close to one. The second periodic window, near $\delta T=4.1$--$5.2$, retains only a narrow stochastic peak of about $0.37$ near $\delta T=4.75$. The third, near $\delta T=7.7$--$7.9$, has no appreciable stochastic success in the sampled realizations. Thus, the principal window is comparatively robust to the sampled switching variability, whereas the additional windows are substantially more sensitive to it.

\begin{figure}[!tbp]
    \centering
    \includegraphics[width=\linewidth]{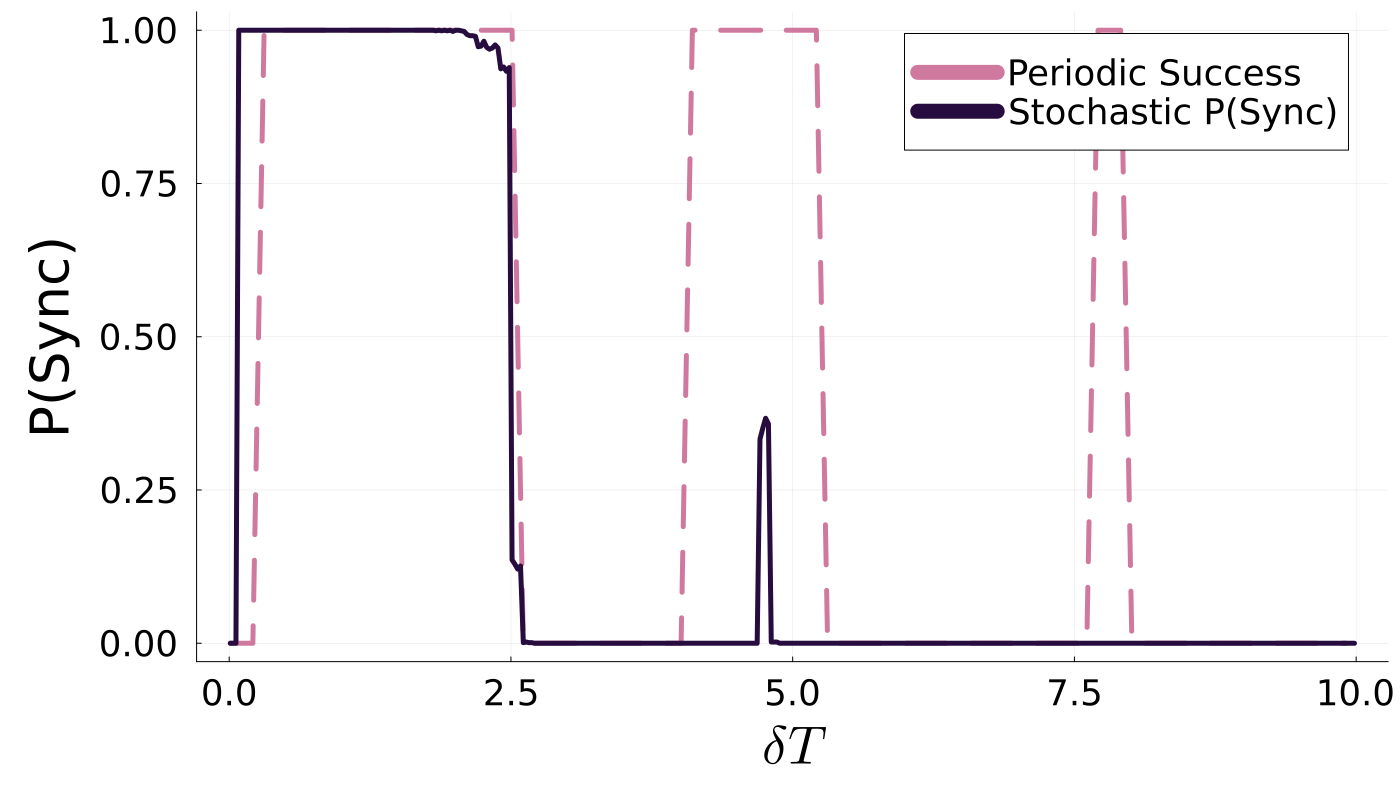}
    \caption{Periodic and stochastic on--off switching in a $50$-node R\"ossler network at $d=3$. {The on-phase topology is a connected Erd\H{o}s--R\'{e}nyi graph with $109$ edges (edge density $\approx0.089$, mean degree $4.36$), $-\gamma_2\simeq0.52$, and $-\gamma_{50}\simeq11.22$.} The pink dashed curve records success or failure of the periodic schedule under the numerical synchronization criterion, and the dark purple curve is the success fraction among $1,000$ independent stochastic realizations, using a tolerance of $10^{-4}$. Both schedules support synchronization over the principal window, approximately $\delta T\in(0.31,2.51)$. A narrow stochastic peak, reaching about $0.37$, remains inside the second periodic window; the third has no appreciable stochastic success. The horizontal axis uses the switching-interval parameter of the simulations.}
    \label{Stochastic vs periodic rossler}
\end{figure}

The piecewise-linear network exhibits the same distinction. Figure~\ref{Stochastic vs periodic piecewise} uses $N=50$, $d=2$, and the same network and switching statistics as the R\"ossler comparison. Periodic switching succeeds over a principal interval $\delta T\in(0.61,1.49)$ and a second interval $\delta T\in(4.41,4.79)$. Stochastic switching supports synchronization across much of the principal interval, with a visible dip in the success probability, but does not reproduce the second periodic window in the sampled runs.

In some intervals, the stochastic success fraction is high while the periodic run does not reach the error tolerance within the observation time. For the piecewise system, the supplied simulations report this behavior near $\delta T\in(0.21,0.58)$ and $(1.51,1.58)$, where the periodic error remains of order $10^{-3}$ at the end of the run. Such a threshold crossing should not be interpreted as proof of a positive asymptotic transverse exponent. The distinction between stability and convergence within a prescribed observation time is particularly relevant near the edges of a synchronization window.

\begin{figure}[!tbp]
    \centering
    \includegraphics[width=\linewidth]{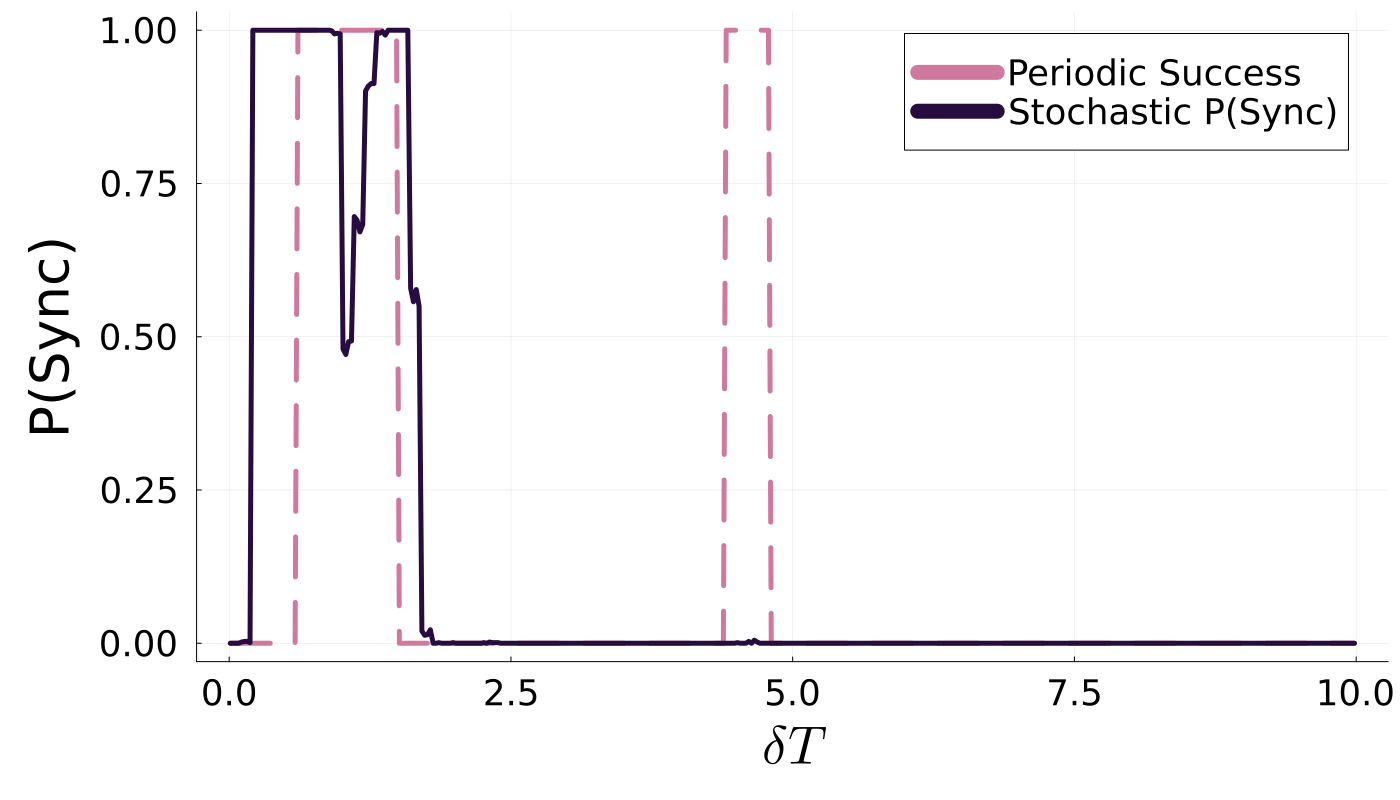}
    \caption{Periodic and stochastic on--off switching in a $50$-node piecewise-linear network at $d=2$, using the same switching statistics and $10^{-4}$ tolerance as Fig.~\ref{Stochastic vs periodic rossler}. The periodic success indicator is pink and dashed; the stochastic success fraction over $1,000$ realizations is dark purple. The principal periodic window is approximately $\delta T\in(0.61,1.49)$, across which stochastic success is substantial but nonuniform. The second periodic window, $\delta T\in(4.41,4.79)$, has essentially no observed stochastic success. Differences near the principal window also reflect the finite observation time and threshold used to classify synchronization.}
    \label{Stochastic vs periodic piecewise}
\end{figure}

The comparison across both models separates two aspects of the switching mechanism. The principal window can survive substantial variation in the sampled switching sequences, consistent with earlier observations of stochastic windows of opportunity. \cite{jeter2015synchronization} The additional periodic windows are more sensitive to this variability: they are strongly suppressed or absent from the stochastic success curves. This difference limits the direct use of periodic stability diagrams for random schedules and motivates a separate analysis of the corresponding random matrix products.

\section{Conclusions}
Analytical descriptions of non-fast switching dynamics remain scarce,
with notable exceptions including antiresonance stabilization and
mechanical oscillators with switched parameters.
\cite{porfiri2021antiresonance,xiedeng2026dynamic} Here, we identify an
analytically tractable setting within a class of saddle-focus
oscillators whose synchronization dynamics can be approximated by their
focal-plane motion. This structural property reduces the synchronization
problem to a linear system with constant coefficients during each
switching phase, yielding explicit Floquet multipliers and stability
conditions. The resulting criteria are exact for the reduced
variational system and provide analytical predictions for the original
networks, as demonstrated through comparisons with R\"ossler and
piecewise-linear oscillators. The reduction thus provides a route to
understanding non-fast switching that is generally unavailable for
continuous-time networks with trajectory-dependent variational dynamics.

The analysis also identifies a geometric origin of the windows of
opportunity previously observed numerically.
\cite{jeter2015synchronization} In the R\"ossler example, rotation
during the uncoupled phase repeatedly reorients transverse
perturbations toward the contracting direction of the coupled saddle.
Suitable switching times exploit this contraction before expansion
along the unstable direction erases the gain. Repetition produces net
decay even when both frozen modes and their average are unstable.
This mechanism gives a concrete realization, within the reduced
synchronization dynamics, of the latent stability suggested by
antiresonance in switched systems.
\cite{porfiri2021antiresonance} The occurrence of related windows in
the piecewise-linear example supports their interpretation as a
consequence of shared focal geometry, rather than a peculiarity of
the R\"ossler equations.

The reduction may extend beyond the two oscillator models considered
here. In both examples, the focal motion is captured by the $(x,y)$
variables, while the nonlinearity is confined to the $z$ equation.
Omitting the corresponding $z$ perturbation from the variational
description therefore leaves a linear focal subsystem with constant
coefficients. Other saddle-focus oscillators may admit an analogous
reduction through a suitable change of variables that separates the
focal dynamics from the remaining direction. Establishing such a
reduction would require assessing the nonlinear feedback into the
focal subsystem and showing that its stabilization remains sufficient
for full transverse stability. Natural candidates include Lur'e
systems with a scalar static nonlinearity, which consist of a linear
dynamical subsystem interconnected with a nonlinear feedback element.
\cite{suykens1999robust} This class also includes saddle-focus Chua-type oscillators.\cite{chua1986double} Identifying the
conditions under which these systems admit an accurate focal reduction
is a promising direction for broadening the present analytical framework.

A central consequence is that switching time and network topology
become complementary design variables. The master stability diagram
connects the switching schedule to the effective couplings and network topology, making it possible to determine whether all transverse
network modes lie in a stable region. At fixed topology, the schedule
can be adjusted to seek a window of opportunity. At fixed schedule,
the spectrum can instead be modified through edge pruning or edge
augmentation, as illustrated by the two network examples. This
suggests a reciprocal design principle: when the structure changes,
retune the timing; when the timing is constrained, reshape the
structure. Such compensation is possible whenever an admissible
choice places every transverse mode within the stable region.
The resulting synchronization is achieved by selecting the network
and its switching schedule, without adding a feedback controller or
altering the individual oscillator dynamics.

The analytical construction is not restricted to switching between
two configurations. For any prescribed finite periodic sequence of
topologies and dwell times, the reduced monodromy matrix is obtained
by multiplying the corresponding phase propagators in temporal order.
When the symmetric Laplacians commute pairwise, they share an
orthonormal eigenbasis, and the analysis retains its decomposition
into two-dimensional transverse modes. Thus, the construction extends to more
general periodic network sequences, although the complexity of the
stability calculation increases.

Stochastic switching presents a further challenge. In both oscillator
examples, the sampled random schedules preserve much of the principal
synchronization window, while the additional periodic windows are
strongly suppressed or absent. These numerical observations show that
periodic stability cannot be transferred directly to random schedules.
An extension to Markovian switching would allow temporal correlations
and switching memory to enter the analysis, connecting the present
continuous-time reduction with earlier results for stochastically
coupled maps. \cite{porfiri2017Memory} Such an extension requires
stability conditions for products of random phase propagators and
would help determine how switching memory and network topology
jointly shape the windows of opportunity.

\begin{acknowledgments}
This work was supported by the
Georgia State University’s Brains \& Behavior Program (R.J., A.B., and I.B.), the
National Science Foundation (USA) under Grant DMS-2510860 (to I.B.) and Grant CMMI-2306824 (to M.P.), and the Aspen Italia Award for Scientific Research and Cooperation (to M.P.).
We dedicate this paper to the memory of our colleague and dear friend, Erik Bollt, whose pioneering contributions to synchronization in switching networks continue to inspire our work.
\end{acknowledgments}

\section*{Data Availability}
The numerical data and custom simulation codes that support the findings of this paper are available from the corresponding author upon reasonable request.

\appendix
\section{Algorithms for Adapting the Network Topology}
\label{Topology adaptation algorithms}
\renewcommand{\thealgorithm}{A\arabic{algorithm}}
\setcounter{algorithm}{0}
The following greedy procedures generate the topology sequences used in Figs.~\ref{Pruning algorithm stability} and \ref{Edge augmenting stability}. All graphs are undirected, simple, and unweighted. We use the signed Laplacian $L=A-D$, whose eigenvalues satisfy $0=\gamma_1\geq\gamma_2\geq\cdots\geq\gamma_N$, and whose spectral radius is $\rho(L)=-\gamma_N$. At each step we record both extreme magnitudes $-\gamma_2$ and $-\gamma_N$. These determine whether the spectrum fits within a single stable interval; a stability set with several components requires checking all transverse eigenvalues.

\subsection{Connectivity-preserving edge pruning}
Algorithm~\ref{Minimal spectral radius edge pruning algorithm} evaluates every edge whose removal leaves the graph connected and selects the deletion with the smallest resulting spectral radius. The procedure stops after at most $K$ removals or when no admissible edge remains. Preserving connectivity avoids an additional zero Laplacian mode, which would retain the unstable uncoupled transverse dynamics in the present models. The rule targets $-\gamma_N$, but does not impose a lower bound on $-\gamma_2$; this is why continued pruning can eventually destroy synchronization.

\begin{algorithm}[!tbp]
\caption{Greedy edge pruning with connectivity preservation}
\label{Minimal spectral radius edge pruning algorithm}
\begin{algorithmic}[1]
\Require Connected graph $\mathcal G_0=(\mathcal V,\mathcal E_0)$; removal limit $K$
\Ensure Graph sequence and extreme Laplacian magnitudes
\State $k\gets0$; record $(-\gamma_2(\mathcal G_0),-\gamma_N(\mathcal G_0))$
\While{$k<K$}
    \State $r_{\rm best}\gets\infty$; $e_{\rm best}\gets\varnothing$
    \For{each edge $e$ of $\mathcal G_k$}
        \State $\mathcal G'\gets\mathcal G_k-e$
        \If{$\mathcal G'$ is connected}
            \State $r\gets\rho(L(\mathcal G'))$
            \If{$r<r_{\rm best}$}
                \State $r_{\rm best}\gets r$; $e_{\rm best}\gets e$
            \EndIf
        \EndIf
    \EndFor
    \If{$e_{\rm best}=\varnothing$}
        \State \textbf{break}
    \EndIf
    \State $\mathcal G_{k+1}\gets\mathcal G_k-e_{\rm best}$; $k\gets k+1$
    \State record $(-\gamma_2(\mathcal G_k),-\gamma_N(\mathcal G_k))$
\EndWhile
\State \Return $\mathcal G_0,\ldots,\mathcal G_k$ and recorded spectra
\end{algorithmic}
\end{algorithm}

\subsection{Edge augmentation}
Algorithm~\ref{Minimal spectral radius edge augmenting algorithm} begins with a path and adds one missing edge at each step, selecting the candidate with the smallest resulting spectral radius. Connectivity is preserved automatically. The algorithm records both endpoints as their magnitudes increase; whether the change promotes synchronization is determined by the selected master stability slice. A path has $N-1$ edges, so the complete sweep contains $\binom{N}{2}-(N-1)$ additions. Both procedures use a fixed candidate ordering to break ties. Neither procedure guarantees a globally optimal graph or the fewest modifications needed for synchronization.

\begin{algorithm}[!tbp]
\caption{Greedy edge augmentation from a path graph}
\label{Minimal spectral radius edge augmenting algorithm}
\begin{algorithmic}[1]
\Require Path graph $\mathcal G_0=(\mathcal V,\mathcal E_0)$ on $N$ nodes
\Ensure Graph sequence to the complete graph and extreme magnitudes
\State $k\gets0$; record $(-\gamma_2(\mathcal G_0),-\gamma_N(\mathcal G_0))$
\While{$\mathcal G_k$ is not complete}
    \State $r_{\rm best}\gets\infty$; $e_{\rm best}\gets\varnothing$
    \For{each missing edge $e=\{i,j\}$ with $i<j$}
        \State $\mathcal G'\gets\mathcal G_k+e$
        \State $r\gets\rho(L(\mathcal G'))$
        \If{$r<r_{\rm best}$}
            \State $r_{\rm best}\gets r$; $e_{\rm best}\gets e$
        \EndIf
    \EndFor
    \State $\mathcal G_{k+1}\gets\mathcal G_k+e_{\rm best}$; $k\gets k+1$
    \State record $(-\gamma_2(\mathcal G_k),-\gamma_N(\mathcal G_k))$
\EndWhile
\State \Return $\mathcal G_0,\ldots,\mathcal G_k$ and recorded spectra
\end{algorithmic}
\end{algorithm}

\section{Focal-Plane Reduction for the Piecewise-Linear Oscillators}
\label{app:piecewise}
The two-dimensional block of $M_{\rm f}$ in \eqref{eq:piecewise_node_matrices} governs expansion and rotation in the focal region, while the third component contracts at rate $2$. For a transverse mode with effective coupling $-d\gamma_l^{s(t)}$, the focal approximation therefore yields \eqref{two node piecewise reduced variational system}. As for the R\"ossler oscillator, this replaces the full trajectory-dependent stability calculation by two constant matrices. Its use for the full network is assessed numerically; linearity within each state-space region does not make the focal approximation exact across the complete trajectory.

\subsection{Numerical comparisons supporting the reduction}
Figure~\ref{piecewise LLE Numerical vs. Analytical} compares the full and reduced transverse stability calculations under static two-node coupling. Here $\gamma_2=-2$, and the reduced matrix is $\left[\begin{smallmatrix}q-2d&-1\\1&q\end{smallmatrix}\right]$. Its stability interval is $q<d<(q^2+1)/(2q)$, or $0.2<d<2.6$ for $q=0.2$. The full calculation has a similar bounded interval. Figure~\ref{piecewise ILE color coded attractors} illustrates the associated instantaneous convergence and divergence along the synchronous trajectory. These diagnostics motivate retaining the focal dynamics, while the periodic comparisons in Fig.~\ref{Analytical stability region for on-off switching in two node system.} directly assess its prediction of switching windows.

\begin{figure}[!tbp]
    \centering
    \includegraphics[width=\linewidth]{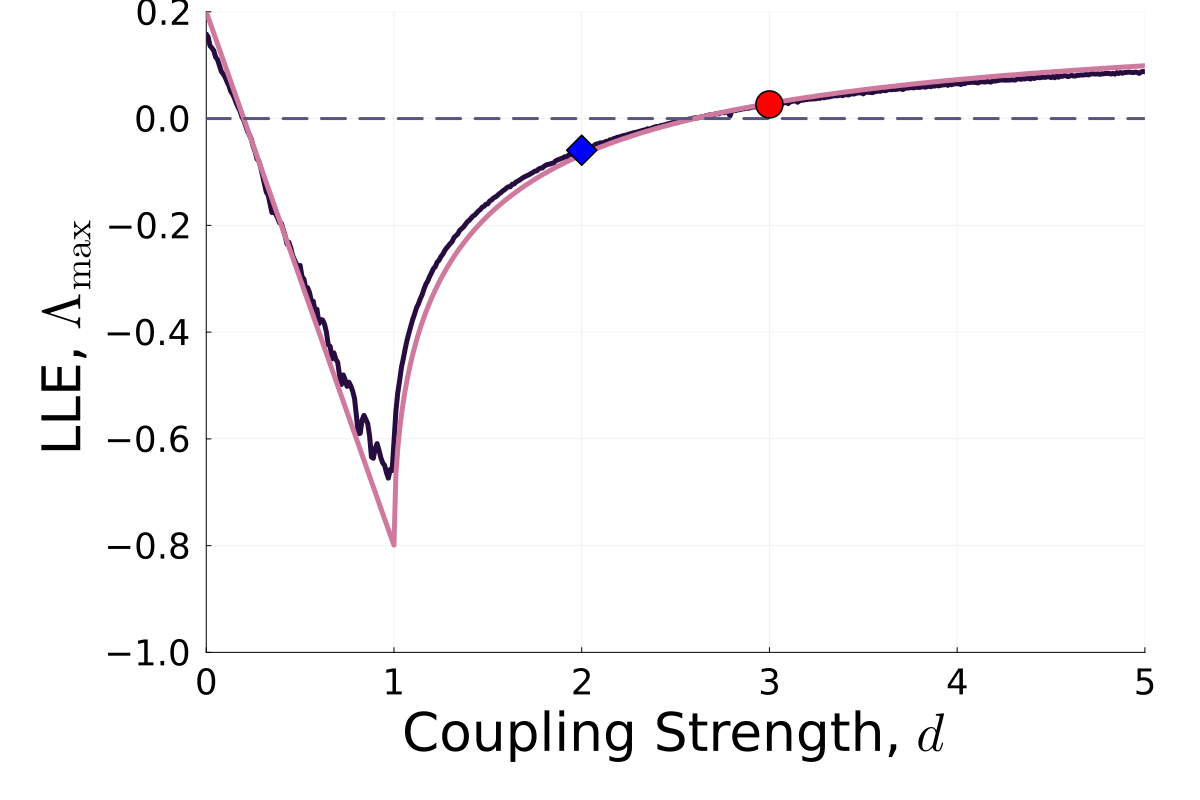}
    \caption{Static two-node stability comparison for the piecewise-linear oscillator. The full-system numerical transverse exponent (dark purple) is compared with the largest real part of the eigenvalues of the focal reduced matrix (pink). Both calculations exhibit a bounded interval of stable synchronization. The blue diamond and red circle identify the couplings $d=2$ and $d=3$ used in Fig.~\ref{piecewise ILE color coded attractors}.}
    \label{piecewise LLE Numerical vs. Analytical}
\end{figure}

\begin{figure}[!tbp]
    \centering
    \includegraphics[width=\linewidth]{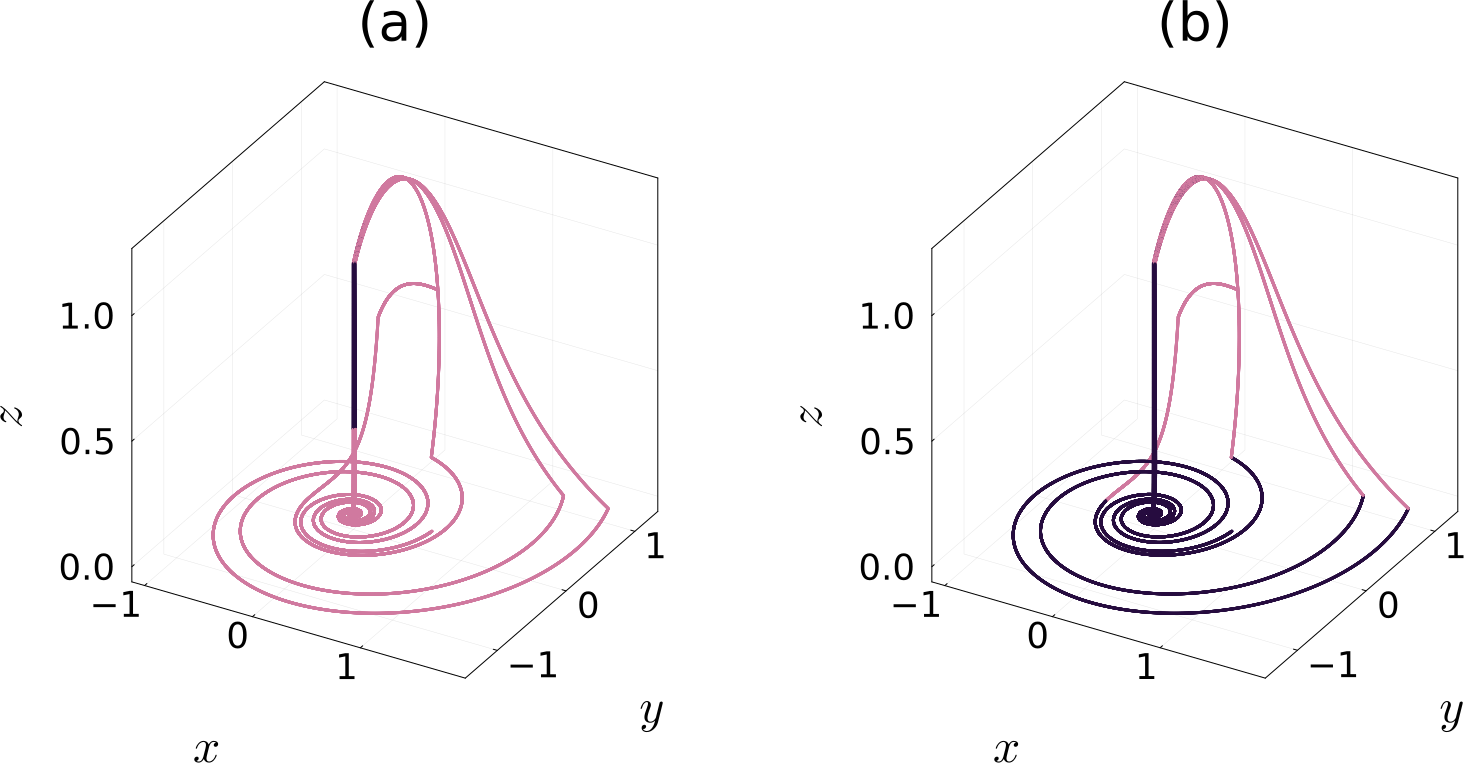}
    \caption{Synchronous attractor of the piecewise-linear oscillator, colored by the numerical instantaneous transverse-stability diagnostic under static coupling: (a) $d=2$ and (b) $d=3$. Dark purple indicates instantaneous divergence and pink indicates instantaneous convergence. The contrasting behavior in the focal region motivates the reduction used to calculate the master stability diagram in Fig.~\ref{Analytical stability region for on-off switching in two node system.}.}
    \label{piecewise ILE color coded attractors}
\end{figure}

\subsection{Monodromy matrix and scalar stability conditions}
{For on--off coupling, the numerical superscripts $0$ and $1$ denote the off and on modes, respectively. The two reduced matrices are}
\begin{equation}
    B_{{\rm pw},l}^{(0)}=\begin{bmatrix}q&-1\\1&q\end{bmatrix},
    \qquad
    B_{{\rm pw},l}^{(1)}=\begin{bmatrix}q+d\gamma_l&-1\\1&q\end{bmatrix}.
    \label{piecewise reduced variational system on-off matrices}
\end{equation}
For arbitrary duty cycle, their monodromy matrix is $\Phi_{{\rm pw},l}=\exp[B_{{\rm pw},l}^{(1)}(1-\delta)T]\exp[B_{{\rm pw},l}^{(0)}\delta T]$, with determinant $\exp\{[2q+(1-\delta)d\gamma_l]T\}$. For the equal dwell times used in the figures, set $\tau=\delta T=T/2$ and define
\begin{equation}
    \begin{aligned}
        k_l^2&=\frac{(d\gamma_l)^2}{4}-1,
        &C_l&=\cosh(k_l\tau),\\
        S_l&=\begin{cases}\sinh(k_l\tau)/k_l,&k_l\ne0,\\\tau,&k_l=0.\end{cases}
    \end{aligned}
    \label{eq:piecewise_phase_constants}
\end{equation}
As in \eqref{eq:phase_functions}, imaginary $k_l$ gives real trigonometric expressions. Direct multiplication gives
\begin{equation}
    \begin{aligned}
        \chi_{{\rm pw},l}
        &=2e^{(2q+d\gamma_l/2)\tau}
            [\cos\tau\,C_l-\sin\tau\,S_l],\\
        D_{{\rm pw},l}&=e^{(4q+d\gamma_l)\tau}.
    \end{aligned}
    \label{eq:piecewise_invariants}
\end{equation}
The quadratic Jury conditions \eqref{eq:jury_monodromy} then reduce to
\begin{equation}
    -d\gamma_l>4q,\qquad
    |\chi_{{\rm pw},l}|<1+D_{{\rm pw},l},
    \label{eq:piecewise_jury}
\end{equation}
for each $l=2,\ldots,N$. The multipliers follow from \eqref{eq:floquet_multipliers} with these trace and determinant, and their maximum modulus defines $\mathcal{L}_{{\rm pw},{\rm red}}=T^{-1}\ln\rho(\Phi_{{\rm pw},l})$. Along the slice $-d\gamma_l=4$, the complex-multiplier portion has exponent $q+d\gamma_l/4=-0.8$, explaining the plateau in Fig.~\ref{Analytical stability region for on-off switching in two node system.}(b).

Evaluating \eqref{eq:piecewise_jury} at $\tau=1.75$ gives the threshold $-d\gamma_l\simeq3.427533$ used in \eqref{augmentation criterion}. The strong-coupling limit provides a further check on this slice:
\begin{equation}
    \lim_{-d\gamma_l\to\infty}\rho(\Phi_{{\rm pw},l})
       =e^{2q\tau}|\cos\tau|\simeq0.359<1.
    \label{eq:piecewise_strong_coupling}
\end{equation}
This contrasts with an upper stability obstruction and explains why the augmentation experiment can continue toward a complete graph without leaving the reduced stable set over the relevant coupling range.
\endgroup

\bibliographystyle{aipnum4-1}
\bibliography{references_new}

\end{document}